\documentstyle[11pt, aas2pp4,psfig,epsf]{article}

\def\plottwo#1#2{\centering \leavevmode
\epsfxsize=.45\textwidth \epsfbox{#1} \hfil
\epsfxsize=.45\textwidth \epsfbox{#2}}

\def\kms{\relax \ifmmode {\, \rm km\, s}^{-1}\else \, km\, s$^{-1}$\fi}
\def\ha{\relax \ifmmode {\rm H}\alpha\else H$\alpha$\fi}
\def\hb{\relax \ifmmode {\rm H}\beta\else H$\beta$\fi}
\def\hi{\relax \ifmmode {\rm H {\sc i}}\else H {\sc i}\fi}  
\def\hii{\relax \ifmmode {\rm H~{\sc ii}}\else H~{\sc ii}\fi}
\def\oiii{\relax \ifmmode {\rm O\, {\sc iii}}\else O\, {\sc iii}\fi}
\def\oii{\relax \ifmmode {\rm O\, {\sc ii}}\else O\, {\sc ii}\fi}
\def\lha{\relax \ifmmode L_{{\rm H}\alpha}\else $L_{{\rm H}\alpha}$\fi}
\def\shi{\relax \ifmmode \ $\sigma$_{{\rm HI}}\else $\ $\sigma$_{\rm HI}$\fi}
\def\sh2{\relax \ifmmode \ $\sigma$_{{\rm H}_2}\else $\ 
$\sigma$_{{\rm H}_2}$\fi}

\def\ie{{\it i.e.},}
\def\eg{{\it e.g.},}

\def\me{$^{-1}$}              
\def\ref#1{{\hangindent=\parindent \hangafter=1 \par \noindent #1}}
\def\degr{\hbox{$^\circ$}}
\def\arcmin{\hbox{$^\prime$}}
\def\arcsec{\hbox{$^{\prime\prime}$}}

\def\fdg{\hbox{$.\!\!^\circ$}}
\def\fs{\hbox{$.\!\!^{\rm s}$}}
\def\farcm{\hbox{$.\mkern-4mu^\prime$}}
\def\farcs{\hbox{$.\!\!^{\prime\prime}$}}
\def\degd#1.#2{ #1\fdg#2 }                 
\def\mind#1.#2{ #1\farcm#2 }               
\def\secd#1.#2{ #1\farcs#2 }               
\def\hhh{\ifmmode {\rm ^h}              
         \else {${\rm ^h}$} 
         \fi}
\def\sss{\ifmmode {\rm ^s}              
         \else {${\rm ^s}$}
         \fi}

\def\hms#1h#2m#3s{                      
                  \relax
                  \ifmmode #1^{\rm h}\, #2^{\rm m}\, #3^{\rm s}
                  \else \hbox{$#1^{\rm h}\, #2^{\rm m}\, #3^{\rm s}$}
                  \fi
                 }
\def\dms#1d#2m#3s{                      
                  \relax
                  #1\degr\, #2\arcmin\, #3\arcsec
                 }
\def\hmsd#1h#2m#3.#4s{                  
                      \relax
                      \ifmmode #1^{\rm h}\, #2^{\rm m}\, #3\fs#4
                      \else \hbox{$#1^{\rm h}\, #2^{\rm m}\, #3\fs#4$}
                      \fi
                     }
\def\apj#1,  {{\it ApJ, ~}{\bf#1},  }
\def\apjlett#1,  {{\it ApJL, ~}{\bf#1},  }
\def\apjsupp#1, {{\it ApJS, ~}{\bf#1},  }
\def\aj#1,  {{\it AJ, ~}{\bf#1},  }
\def\astrf#1,  {{\it Astrofizika, ~}{\bf#1},  }
\def\aasupp#1,  {{\it AAS, ~}{\bf#1},  }
\def\aa#1,  {{\it AA, ~}{\bf#1},  }

\def\mnras#1,  {{\it MNRAS, ~}{\bf#1},  }
\def\mmnras#1,  {{\it MemRAS, ~}{\bf#1},  }
\def\annrev#1,  {{\it ARAA, ~}{\bf#1},  }
\def\ass#1,  {{\it Astrophys.\ Space\ Sci.~}{\bf#1},  }
\def\pasp#1,  {{\it PASP, ~}{\bf#1},  }
\def\pasa#1,  {{\it PASA, ~}{\bf#1},  }

\accepted{Jan. 2000 for AJ}

\lefthead{Beckman et al.}
\righthead{Populations of high-luminosity density-bounded HII regions}

\begin{document}

\title{Populations of high-luminosity density-bounded\\ 
H II regions in spiral galaxies?\\ 
Evidence and implications}

\author{J. E. Beckman\altaffilmark{1,2}, M. Rozas\altaffilmark{1},  
A. Zurita\altaffilmark{1}, R. A. Watson\altaffilmark{1}}
\altaffiltext{1}{Instituto de Astrof\'\i sica de Canarias,  E-38200 La 
Laguna, Tenerife,  Spain}
\altaffiltext{2}{Consejo Superior de Investigaciones Cient\'\i ficas, 
CSIC, Spain}

\and

\author{J. H. Knapen\altaffilmark{3,4}}
\altaffiltext{3}{University of Hertfordshire,  Department of Physical 
Sciences, Hatfield,  Herts AL10 9AB,  UK.}
\altaffiltext{4}{On leave at Isaac Newton Group of Telescopes, E-38700
Santa Cruz de La Palma, Spain}

\begin{abstract}

In this paper we present evidence that the \hii\ regions of high
luminosity in disk galaxies may be density bounded, so that a
significant fraction of the ionizing photons emitted by their exciting
OB stars escape from the regions.

The key piece of evidence is the presence, in the \ha\ luminosity
functions (LFs) of the populations of \hii\ regions, of glitches, local
sharp peaks at an apparently invariant luminosity, defined as the
Stromgren luminosity ($L_{\rm Str}$), $L_{{\rm H}\alpha}$ = $L_{\rm
Str}$ = 10$^{38.6}$ ($\pm$ 10$^{0.1}$) erg s\me\ (no other peaks are
found in any of the LFs) accompanying a steepening of slope for $L_{{\rm
H}\alpha}$$>$ $L_{\rm Str}$. This behavior is readily explicable via a
physical model whose basic premises are: (a) the transition at $L_{{\rm
H}\alpha}$ = $L_{\rm Str}$ marks a change from essentially ionization
bounding at low luminosities to density bounding at higher values, (b)
for this to occur the law relating stellar mass in massive star-forming
clouds to the mass of the placental cloud must be such that the ionizing
photon flux produced within the cloud is a function which rises more
steeply than the mass of the cloud. Supporting evidence for the
hypothesis of this transition is also presented: measurements of the
central surface brightnesses of \hii\ regions for $L_{{\rm H}\alpha}$
$<$ $L_{\rm Str}$ are proportional to $L_{{\rm H}\alpha}^{1/3}$,
expected for ionization bounding, but show a sharp trend to a steeper
dependence for $L_{{\rm H}\alpha}$ $>$ $L_{\rm Str}$, and the observed
relation between the internal turbulence velocity parameter, sigma, and
the luminosity, $L$, at high luminosities, can be well explained if
these regions are density bounded.
	
If confirmed, the density-bounding hypothesis would have a number of
interesting implications. It would imply that the density-bounded
regions were the main sources of the photons which ionize the diffuse
gas in disk galaxies. Our estimates, based on the hypothesis, indicate
that these regions emit sufficient Lyman continuum not only to ionize
the diffuse medium, but to cause a typical spiral to emit significant
ionizing flux into the intergalactic medium. The low scatter observed in
$L_{\rm Str}$, less than 0.1 mag rms in the still quite small sample
measured to date, is an invitation to widen the data base, and to
calibrate against primary standards, with the aim of obtaining a
precise, $\sim 10^5$ L$_{\odot}$, widely distributed standard candle.

\end{abstract}

\keywords{ISM: \hii\ regions -- galaxies: individual (NGC~157, NGC~3631,
NGC~6764, NGC~6951, NGC~6814, NGC~5194, NGC~7479, NGC~3359) -- galaxies:
spiral}

\vfill{Accepted for publication in the Astronomical Journal}

\section{Introduction}

The basic theory for modeling a gaseous region round a hot star was
first given by Zanstra (1931) for planetary nebulae, and applied to
\hii\ regions by Str\"{o}mgren (1939), who quantified the relation
between the radius of the ionized zone and the temperature--luminosity
of the central exciting star. He showed that, in a uniform medium, the
transition layer between fully ionized and neutral gas will be thin
compared to the radius, a structure since termed a Str\"{o}mgren
sphere. This will occur in a large placental neutral cloud, of
sufficient dimension to absorb all the ionizing photons (those in the
Lyman continuum in the case of atomic hydrogen), in which case the \hii\
region is ``ionization bounded". The case where the cloud is not big
enough to absorb all the ionizing radiation was first dealt with by
Hummer \& Seaton (1964) for planetary nebulae, which are then termed
optically thin. An \hii\ region formed in these circumstances, where the
cloud radius is less than that of the Str\"{o}mgren sphere, is termed
``density bounded" or ``matter bounded".

There is ample evidence that \hii\ regions are not homogeneous in
density. Measurements of local electron densities via line ratios are
typically two orders of magnitude higher than mean electron densities
estimated using diametral emission measure (for a clear recent example
see Rozas, Knapen, \& Beckman 1996), explained if a region comprises
knots of high density embedded in a lower density plasma. The fractional
volume dense enough to contribute measurably in emission lines such as
\ha\ has been termed the ``filling factor" ({\it e.g.}, Osterbrock
1989). An \hii\ region with this structure will form a Str\"{o}mgren
sphere provided that the local density variations are on scales which
are small compared with the \hii\ region diameter, and that the mean
density varies little on the scale of the diameter. This condition
appears to hold well for regions over a wide range of \ha\ luminosities,
up to a critical value, which we will be able to quantify from
observations, as described below.  The regions in this range are
ionization bounded. At higher luminosities our evidence, which will be
described in this paper, points towards an increasing tendency for the
photon output from the ionizing stars to overflow the cloud in which
they have formed.

One type of evidence comes from the measurements of a change in slope of
the luminosity function in \ha\ of complete populations of \hii\ regions
in the set of nearby spirals. Previous detections of this change have
been reported in the literature, notably in earlier work by Hodge, and
by Kennicutt and co-workers (Hodge 1987; Kennicutt 1984; Kennicutt,
Edgar, \& Hodge 1989); and most recently by McCall and co-workers
(McCall, Straker, \& Uomoto 1996; Kingsburgh \& McCall 1998) as well as
ourselves (Rozas, Beckman, \& Knapen 1996). It is interesting to note,
however, that the change in slope varies little from object to
object. We will elaborate this point during the development of the
present paper.

Statistical studies of the relation between the \ha\ luminosities,
$L_{{\rm H}\alpha}$, of complete samples of \hii\ regions in nearby
large spirals, and their volumes, (Cepa \& Beckman 1989, 1990; Knapen
{\it et al.} 1993; Rozas, Beckman, \& Knapen 1996), show that to a first
approximation these are proportional. Taken as a complete description of
the observations this would imply, at least statistically, two
properties: (1) The \hii\ regions obey a single physical regime, which
should be that of ionization bounding, and (2) the densities of the
clouds in which the \hii\ regions are formed differ little from cloud to
cloud within a galaxy, and from galaxy to galaxy. We will see that at
very high luminosities departure from this first order behavior is
dominant.  Illustration of these points may be found in a number of
articles in the literature, where the \ha\ luminosities of individually
measured \hii\ regions have been plotted, logarithmically, {\it vs.} the
cubes of their radii (Cepa \& Beckman 1989, 1990; Knapen {\it et al.}
1993; Rozas Knapen, \& Beckman 1996; Rozas {\it et al.} 1999).  In these
articles, it is shown that for eight galaxies there is a close to linear
relation between the \ha\ luminosity and the volume of the \hii\ regions
in the range 37.5 $<$ log $L_{{\rm H}\alpha}$ $<$ 39.5, as pointed out
in Cepa \& Beckman (1990).  This implies that the product of the filling
factor, the electron density and the ionized hydrogen density does not
change very much over this range, although in all the galaxies there is
a clear tendency for it to increase at the highest luminosities.

It is not entirely ruled out that the filling factor and mean density
might vary independently with a relation which leaves their product
invariant, but this would amount to a ``conspiracy", and it is more
reasonable to assume that none of the cited parameters varies
strongly. The general picture is of \hii\ regions as spongy structures,
which may indeed have fractal characteristics as suggested by Elmegreen
(1997), but with a degree of clumping, and an average density which do
not depend strongly on luminosity. However in all the plots of log
$L_{{\rm H}\alpha}$ {\it vs.} $r^3$ cited above, there is a significant
trend to steeper gradients at higher luminosities, which means that here
the densities and/or the filling factors of the regions must be
increasing.

The evidence which we present below points to a rather sharp change in
the properties of the \hii\ regions occurring at an \ha\ luminosity
which appears to vary very little from galaxy to galaxy. We will show
that one would in fact expect, on reasonable physical grounds, the most
luminous regions to be producing more ionizing photons than can be
absorbed in their placental clouds, whereas this should not be true
globally for the less luminous regions.  We can predict, from a model in
which there is a transition between ionization bounding at lower
luminosities, and density bounding at higher luminosities, that there
should be an accumulation of \hii\ regions around the luminosity of the
transition (which we have termed $L_{\rm Str}$, \ie\ the Str\"{o}mgren
transition), between the two regimes, and this accumulation is in fact
found as a ``glitch'' in the \ha\ luminosity function of the galaxies
observed. It is by no means easy to reproduce these observations with
alternative hypotheses. Two aspects of the density-bounding phenomenon
would be of particular interest. Since the most luminous regions are
those from which we predict the highest proportion of Lyman continuum
(Lyc) photons are escaping, these regions are candidates for the
principal ionizing sources of the diffuse insterstellar medium in large
spirals, and may contribute significantly to the intergalactic ionizing
field. Also the transition between ionization bounding and density
bounding appears to take place over a narrow range in $L_{{\rm
H}\alpha}$, and while the sample of results presented here is not large,
this point is worth following up as a high-luminosity feature of this
kind would have its uses as a powerful, precise, and non-transient
standard candle. However the detailed interpretation of the surface
brightness data is not obvious and further, spectroscopic, observations
will be important to follow up the arguments presented here.

In Sect. 2 we present the evidence about the change in properties of the
\hii\ regions, in Sect. 3 we present scaling relations which support the
model of density bounding for the highest-luminosity regions, and a set
of alternative models aimed at explaining the observed ``glitches" in
the luminosity functions of \hii\ regions in the galaxies observed. In
Sect. 4 we discuss the implications of our models for the diffuse \ha\
emission from the warm ionized medium. In Sect. 5 we summarize our
conclusions, outline the observations required to deepen our
understanding of the transition phenomenon, and suggest applications to
the correction of the global star formation rates (SFRs) in galaxies,
and to the measurement of intergalactic distances.

\section{The evidence for a change in physical regime}

\subsection{The luminosity functions}

	
\begin{figure*}
\plottwo{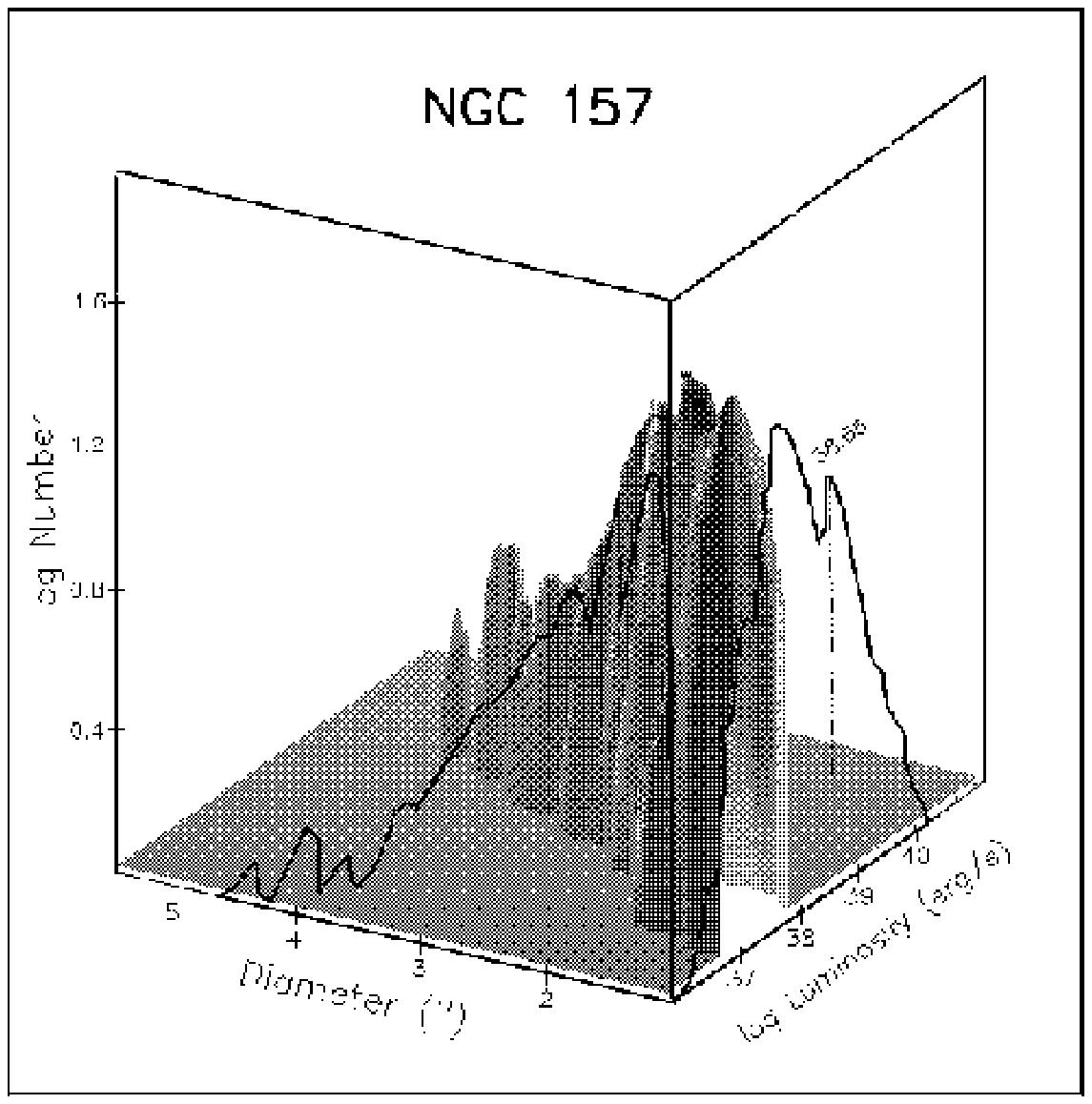}{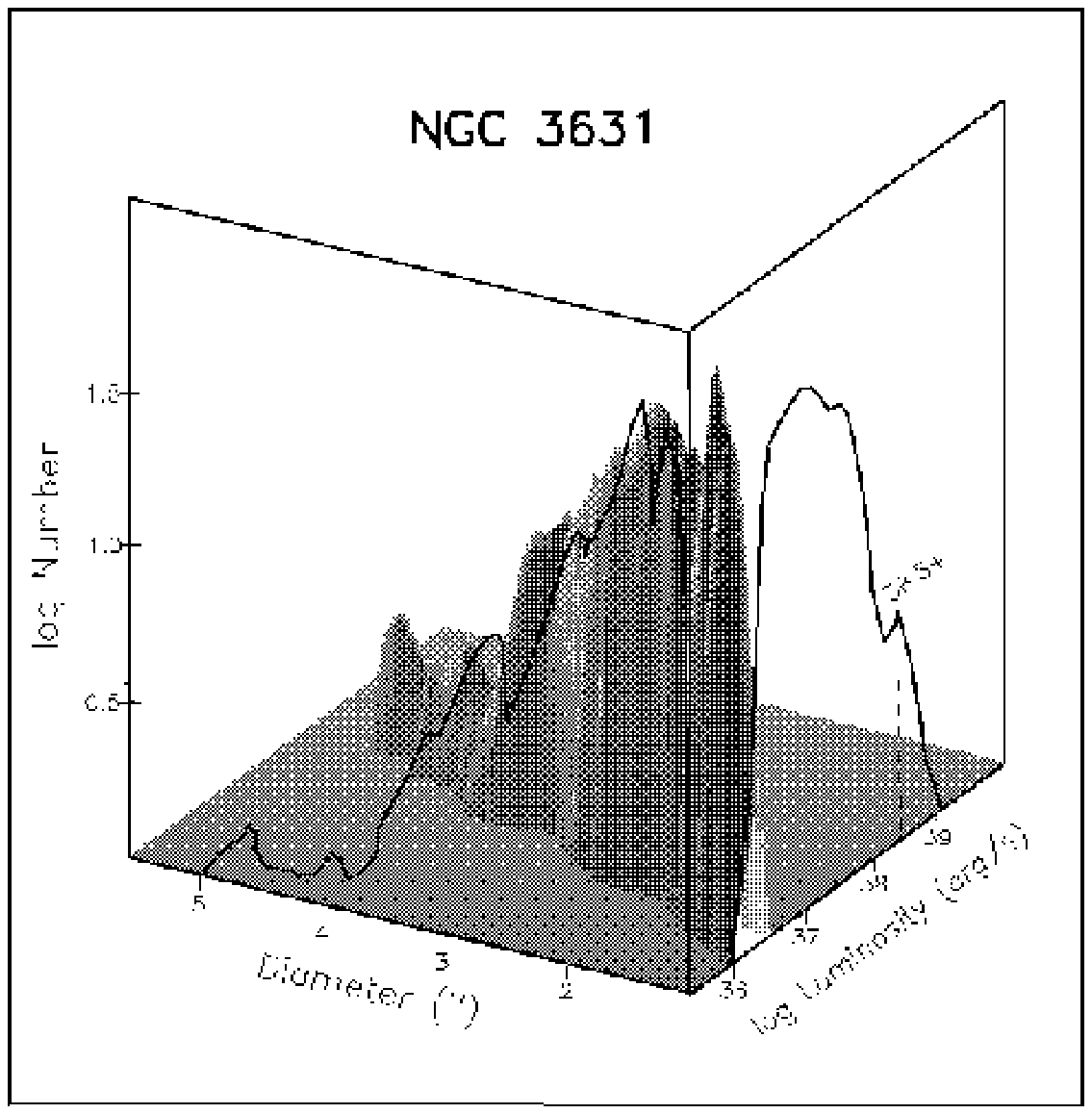}
\plottwo{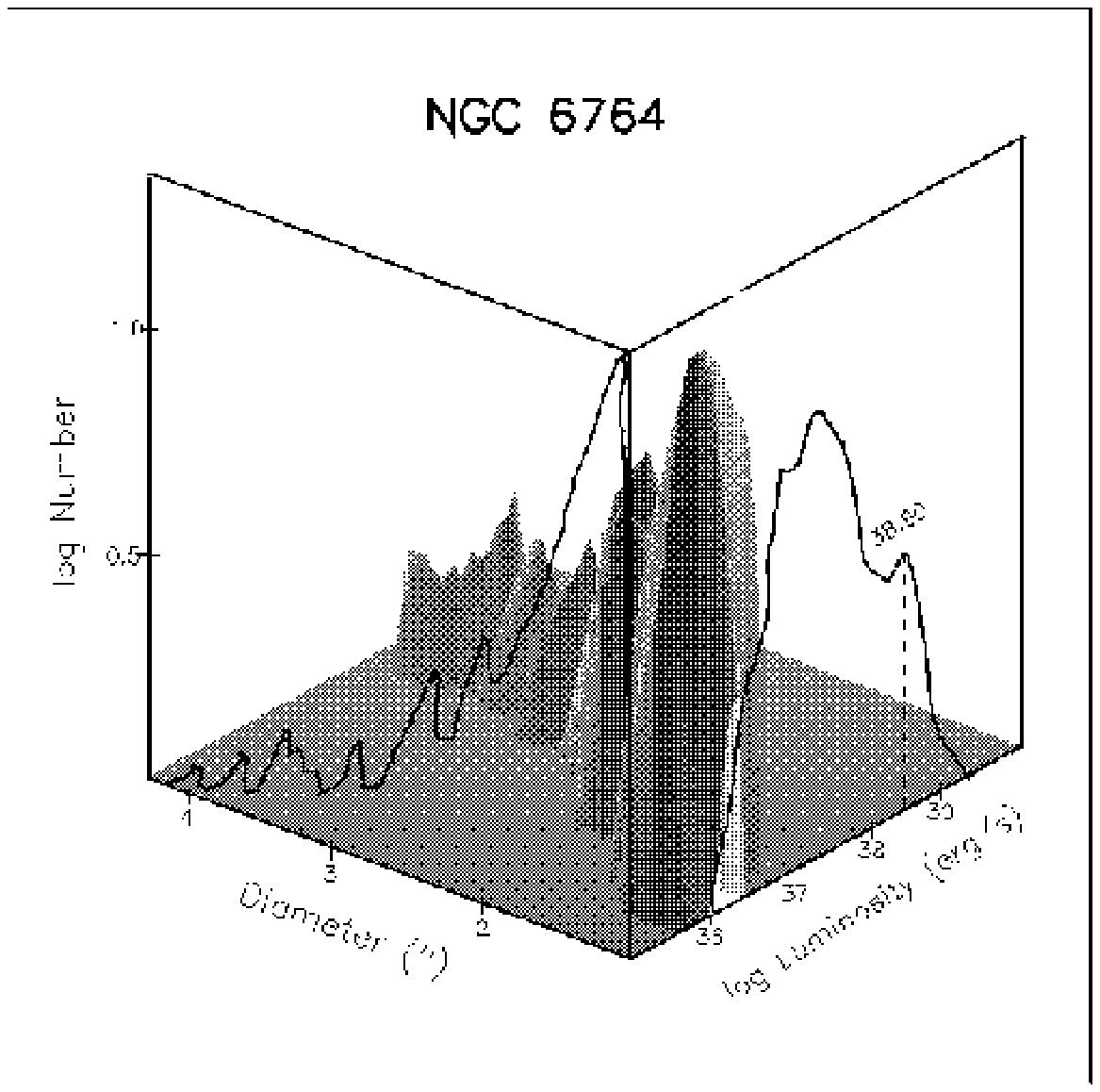}{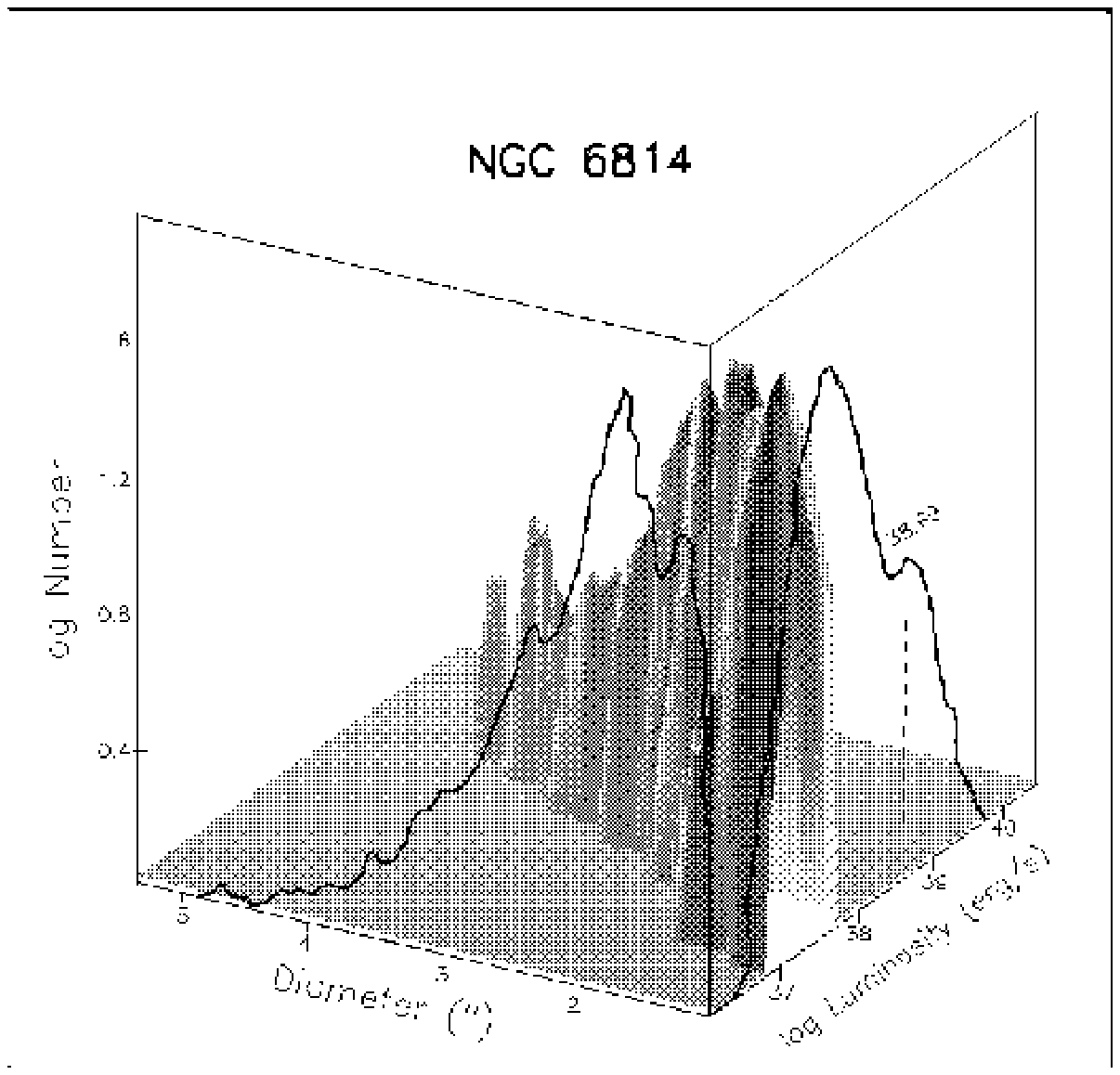}
\caption{Three-dimensional surfaces: best fits to the observed data
representing the numbers, $N$, diameters, $d$, and H$\alpha$
luminosities, $L_{{\rm H}\alpha}$, of the \hii\ regions in six galaxies.
The projections in the log $N$--log $L_{{\rm H}\alpha}$ plane are
luminosity functions, LFs with effective binning of 0.1 dex. The broad
peaks and declines to low $L_{{\rm H}\alpha}$ are caused by
observational selection: low-luminosity regions are increasingly
difficult to measure reliably below log $L_{{\rm H}\alpha}$ $\sim$ 37,
though all samples are complete to below log $L_{{\rm H}\alpha}$ = 38.
The sharp peaks at $L_{{\rm H}\alpha}$ $\sim$ 38.6 are presumably due to
the transition from ionization bounding to density bounding (see text
for details).}
\end{figure*}

\setcounter{figure}{0}
\begin{figure*}
\plottwo{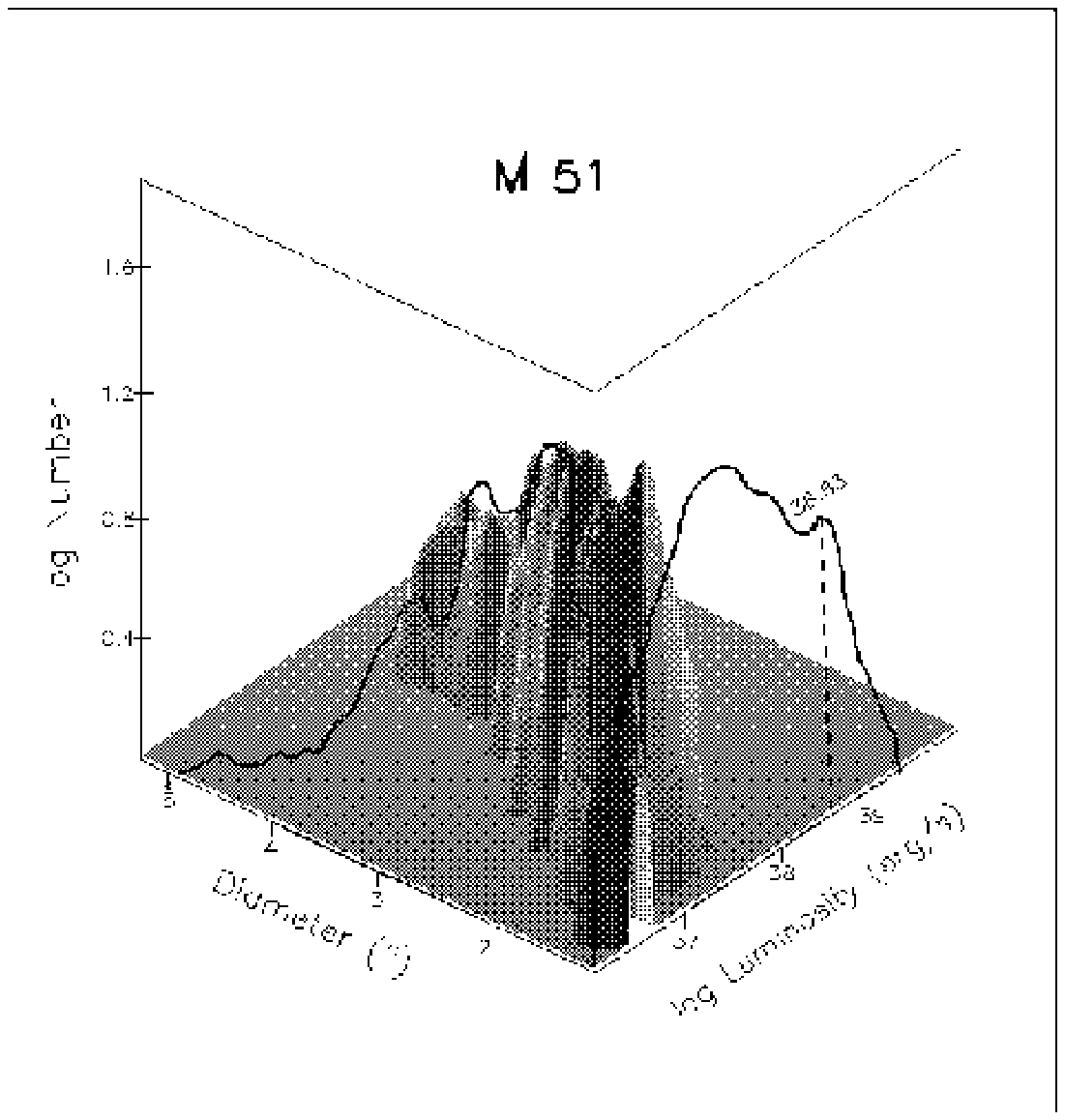}{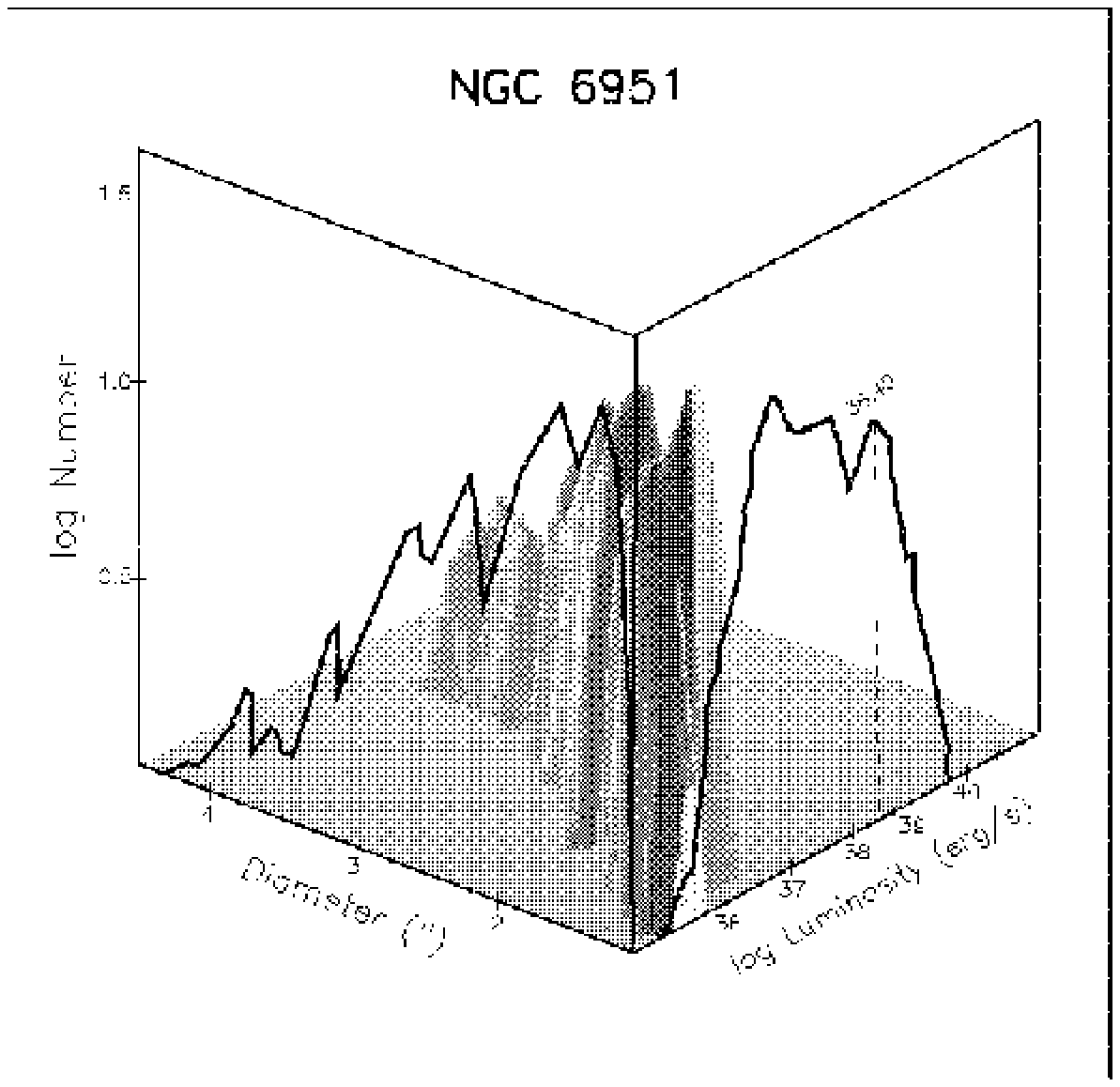}
\caption{Continued}
\end{figure*}

In a set of published papers stretching back over the past decade we
have published luminosity functions (LFs) in \ha\ for the \hii\ regions
in spiral galaxies, the majority of which were in absolutely calibrated
fluxes (Cepa \& Beckman 1989, 1990; Knapen {\it et al.\/} 1993; Rozas,
Beckman, \& Knapen 1996). It was while inspecting these LFs, with the
addition of the LF from Rand (1992), that we found the feature which led
us to postulate the transition from ionization bounding to density
bounding. All the LFs showed an abrupt change in slope accompanied by a
``glitch'': a jump upwards in the value of the LF, at a luminosity close
to log $L_{{\rm H}\alpha}$ = 38.6 ($L_{{\rm H}\alpha}$ in
ergs$^{-1}$). In Fig.~1 we show an improved analysis of the LFs
originally given in Rozas, Knapen, \& Beckman 1996, for the galaxies NGC
157, NGC 3631, NGC 6764, and NGC 6951, in Knapen {\it et al.\/} (1993)
for NGC 6814, and in Rand (1992) for M51, in which the principal change
has been an effective rebinning into bins of 0.1 dex in luminosity. The
new treatment of the data uses a three- dimensional surface fit of log
$L_{{\rm H}\alpha}$ against number count (log $N$) and diameter, which
in Fig.~1 has been projected into the log $L_{{\rm H}\alpha}$--log $N$
plane with a resolution of 0.1 dex. The data show an absence of peaks in
each LF except for the peak at the transition luminosity which, as seen,
is accompanied by a change in the slope of the LF to steeper values at
luminosities higher than that of the transition.

We carried out a specific check on NGC~157 to see how the glitch peak is
affected if instead of using a limiting surface brightness contour to
define the integrated brightnes of each \hii\ region we applied a
cut-off at a constant fraction of the peak luminosity. This technique
(``PPF'') was described in Kinsburgh \& McCall (1996); it avoids
problems of S:N ratios at the boundaries of the \hii\ regions, and
should be superior to the limiting surface brightness method (``FTP'')
for sets of geometrically homologous regions.  If there is a trend to
varying internal brightness gradient with luminosity, the PPF technique
is liable to imprecision. In the present context the most luminous
regions have very bright centres and extended haloes, so that PPF tends
to underestimate their luminosities. This was confirmed with the LF of
NGC~157 which fell off more rapidly at high L$_{H\alpha}$ with PPF than
using our normal technique, i.e. FTP. In spite of this, the glitch
showed up, at the same luminosity but at a statistically less
significant level, due to the smaller numbers of \hii\ regions involved.

To check that a peak is plausibly due to a truly collective property of
the \hii\ regions, and is not merely a statistical fluctuation, we took
the mean value, $N(m)$, of the number of \hii\ regions in the two bins
to either side of the peak, subtracted this from the value $N(max)$ of
the number of regions at the peak, and computed $(N(max) - N(m))/
N(max)^{1/2}$. Values between 1 and 2 were found for each of the
galaxies, with a mean of 1.7. Thus each individual peak is just
significant, but if we were looking at a single galaxy and had no other
kinds of observational evidence we would not, perhaps, want to draw any
striking conclusions from the presence of a peak no higher than
2$\sigma$. However the peaks occur in a number of galaxies, and the
chance that this occurs as the only such feature, at closely similar
luminosity, for purely statistical reasons, is very small, less than 1
part in 10$^3$, so it would be very hard to dismiss the peaks as
statistical artifacts.  The fact that the peaks and gradient changes
occur at the same luminosity within 0.07 mag, for galaxies whose
distances cover a range of over a factor three, also argues strongly
against an observational effect ({\it e.g.}, limitation in angular
resolution) as their cause.

A qualitative explanation for the presence of the glitches in the LFs in
terms of the transition from ionization bounding to density bounding is
as follows.  In a hypothetical set of star-forming clouds with identical
cloud masses but with star clusters of different integrated ionizing
luminosities forming inside them, some of the clusters may be massive
enough to emit more than sufficient ionizing photons than are required
to fully ionize their placental clouds.  In that case, whatever the
intrinsic LF of the resulting \hii\ regions (measured in ionizing
photons) might be, the LF observed in \ha\ would show a spike at the
upper bound of the observed luminosity range, because any regions with
extra ionizing photons would lose the excess by density bounding.  In
this extreme, and not of course realistic, case we would not find a
glitch, but a delta function in the LF.  Cloud masses are not, however,
identical.  In order to yield a glitch two entirely plausible physical
conditions are required: one is that the mass of massive stars rises,
statistically, with the cloud mass; the other is that this rise is
sufficiently steep that at a critical cloud mass, and above it, there
are sufficient Lyc photons to ionize the whole cloud.  We will see in
Sect.  3 that the evidence which does exist in the literature is fully
consistent with these two conditions. 
 
\subsubsection{Modeling of the LFs}

Here we show, using a set of fairly schematic models, that the process
of density bounding ought to give rise to the form of the LF observed in
\ha, and that the other hypothetical causes modeled for the observed
behavior are in no way as plausible.  The model starts by generating an
artificial LF using a population of 1000 \hii\ regions, distributed in
bins of uniform logarithmic width; we chose 0.2 dex for the
illustrations here as this is our preferred plotting scale for the
observations.  The number of regions per bin, $n$, is given by

\begin{equation}
\frac{dN}{d{ \log L}_{{\rm H}\alpha}}=n(L_{{\rm H}\alpha}) = 
\frac{L_{{\rm H}\alpha}^{\xi}}{ A'},
\end{equation}

where $L_{{\rm H}\alpha}$ is the luminosity in \ha, and the index $\xi$
can be varied, but in our figures we will keep this at an
observationally typical value of --1.5.  The constant $A'$ is adjusted
in order to normalize the number of regions at log $L_{{\rm H}\alpha}$ =
38 erg s$^{-1}$ in the 0.2 dex bin to 100.  The form of equation (1) is
the most direct way to represent the luminosity function, but
conventionally either differential or logarithmic forms are used for
convenience.  These are respectively:

\begin{equation}
dN({L}_{{\rm H}\alpha}) = \frac{{L}_{{\rm H}\alpha}^{-\alpha}}{A} \, 
d{L}_{{\rm H}\alpha}
\end{equation}
\vspace{-0.2cm}
and
\vspace{-0.2cm}
\begin{equation}
\frac{dN}{d{\log L}_{{\rm H}\alpha}}  = 
\frac{{L}_{{\rm H}\alpha}^{1-\alpha}}{A'}, 
\end{equation}

where $\alpha=1-\xi$ and $A={A'}/{\rm log}e$.  The form (3) will be used
in the development of Section 3. 

\begin{figure}[h]
\psfig{figure=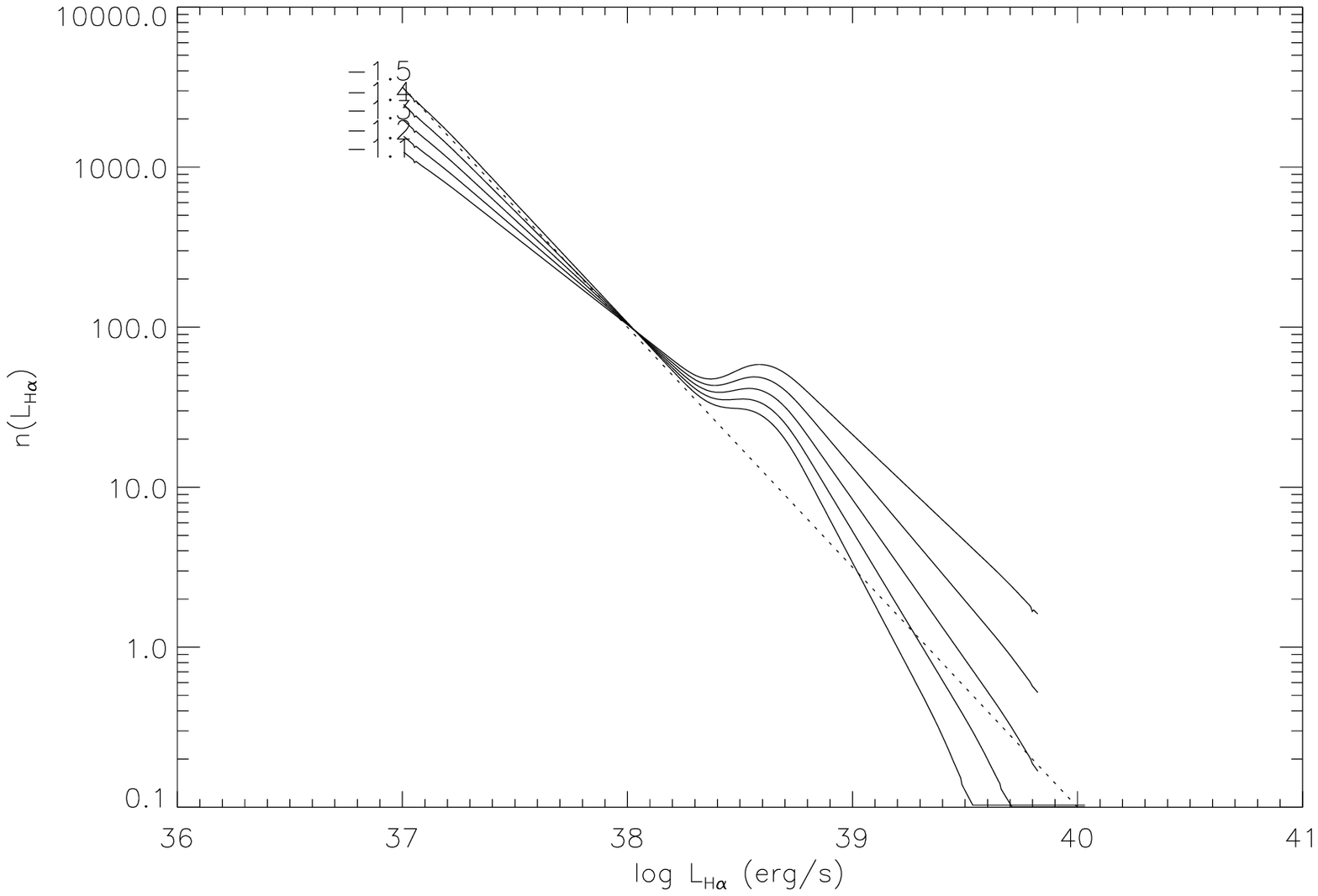,width=8cm}
\caption{Computed models showing how the phenomenon of density
bounding above a critical luminosity can lead to the observed glitch
plus a change of slope in the luminosity functions of the \hii\ regions
of a disk galaxy (see text for details).  Each contour plot corresponds
to an intrinsic slope of the initial LF (values are shown).  The dashed
plot show an LF with slope --1.5 prior to modeling the effect of density
bounding.}
\end{figure}

With the basic function of equation (1), we can then apply processes
which simulate the different physical effects expected to affect the LF. 
In the first place we show a simulation of the effects of density
bounding.  The model assumes that the observed luminosity of a region in
H$\alpha$ is reduced to $L_{H\alpha}$, above the critical value of
$L_{{\rm H}\alpha}$ ($L_{\rm Str}$) at which an \hii\ region is just
fully ionized, by a factor which depends on the ratio of the true
ionizing luminosity (L$_i$) to the ionizing luminosity which produces
$L_{{\rm H}\alpha}$ = $L_{\rm Str}$ = $L_{\rm i,Str}$.  This is a fair
parametric representation of what is predicted when the ionizing flux
overflows the \hii\ region.  The formula for this is just:

\begin{equation}
L_{{\rm H}\alpha} = L'_{{\rm H}\alpha} \times ({\frac {L_{\rm i}}{L_{\rm i,Str}}})^k,
\end{equation}

where $k <$ 0 and $L'_{{\rm H}\alpha}$ is the \hii\ region luminosity in
\ha\ if all the ionizing photons are absorbed.  Given the dependence of
the observed H$\alpha$ luminosity of a region on its intrinsic ionizing
luminosity, we can transform numerically to obtain the corresponding LF
in H$\alpha$.  Applying this to the part of the LF, given by equation
(1) in the range $L_{{\rm H}\alpha}$ $>$ $L_{\rm Str}$ (for $L_{{\rm
H}\alpha}$ $<$ $L_{\rm Str}$, $L_{{\rm H}\alpha}$ = L$'_{H\alpha}$),
with a smoothing function of Gaussian type with $\sigma$ = 0.1 dex in
luminosity (to correspond to the binning of the observational data) we
obtain curves of the type shown in Figure 2.  We can see here that the
general functional form of our observations is reproduced, notably the
glitch in the LF as well as the increase in (negative) slope for
$L_{{\rm H}\alpha}$ $>$ $L_{\rm Str}$.  We can also see that the
luminosity of $L_{\rm Str}$ is almost independent of the slope, $\xi$,
of the initial trial function, which represents the intrinsic slope of
the LF and reflects essentially the IMF slope of the massive exciting
stars within the region, \ie\ the glitch peak luminosity should depend
little on the initial mass function (IMF) if the break is due to density
bounding as we suggest.

To examine another possible cause of the break and the glitch in the LF
we assumed that the \hii\ regions contain an admixture of dust, causing
extinction in H$\alpha$.  Dust is a well-explored feature of the
insterstellar medium, and the main doubt is to what extent it survives
within the extreme conditions inside an ultra-luminous \hii\ region. 
For the purpose of this exercise we assumed a constant dust-to-gas
ratio, and uniform mixing, which is the case most favorable to a dust
extinction hypothesis as the cause of the LF break.  This was treated
via a modification of the intrinsic H$\alpha$ luminosity $L'_{{\rm
H}\alpha}$ given by eq.~(1) ($n(L'_{{\rm H}\alpha})$ = $L'{_{{\rm
H}\alpha}^{\xi}}/A$), to a resultant observed luminosity ${L}_{{\rm
H}\alpha}$, according to the expression:

\begin{equation}
L_{{\rm H}\alpha} =  {L'_{{\rm H}\alpha}}^{-\tau},
\end{equation}

where $\tau$ is an optical depth in H$\alpha$ given by $\tau = B\,r$, $V
= 4/3 \pi r^3$ and the constant $B$ takes values which normalize the
extinction as follows: at a radius of $r$ = 100 pc a region has an
intrinsic luminosity $L'_{{\rm H}\alpha}$ = 10$^{38}$ erg s$^{-1}$
(chosen as a representative value from the observations), which is
proportional to its volume, $V$ (again a dependence taken from the
observations, to first order), and $B$ is then varied at will to give
values of $\tau$ in simple decimal intervals. 

\begin{figure}[h]
\psfig{figure=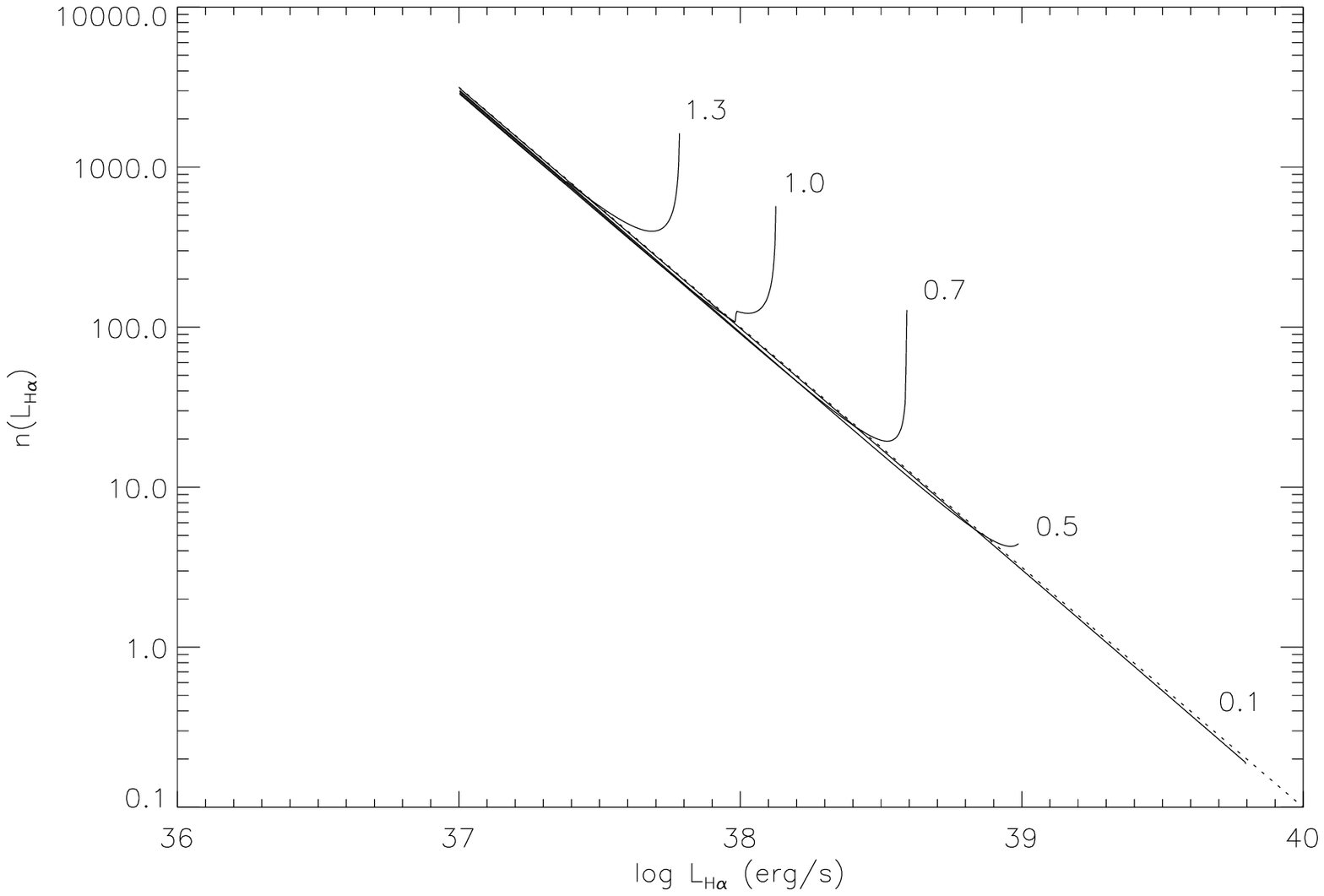,width=8cm}
\caption{Computed models of \hii\ region LFs with constant
gas-to-dust ratio.  Each continuous curve corresponds to a value of
integrated dust optical depth (at the \ha\ wavelength) for a region with
emitted \ha\ luminosity of log~$L_{{\rm H}\alpha}$ = 38 (erg s\me).  The
dotted curve is the unextinguished LF.}
\end{figure}

In Figure 3 we can see the results of this modeling exercise for
different values of the $\tau$ at log $L_{{\rm H}\alpha}$ = 38 erg
s$^{-1}$.  The results show qualitatively very different LFs from what
we observe.  If the optical depth is small the effect is to curtail the
LF but not to alter its slope.  If the optical depth is larger, the
effect is to curtail the LF with a high peak and a drop to zero at a
cut-off luminosity, without altering the slope of the main LF.

Although the result of this exercise with dust suggests that dust
extinction is unlikely to be the cause of the LF we observe, it does
point to a possibility which at least qualitatively might produce an LF
not unlike that observed.  If dust is mixed within the \hii\ regions up
to some limiting luminosity, but then in those of higher luminosity is
expelled, it may be possible to modify the curves in Fig.~2 in such a
way as to displace some of the \hii\ regions accumulated in the rising
tail of the LFs in Fig.~3 to higher luminosities, yielding an LF of the
form observed.  While this scenario is not ruled out physically, it
would give rise to a dependence of the central H$\alpha$ surface
brightness of the \hii\ regions on radius which is very different from
that observed, as we will explain below, and this is sufficient reason
for us not to accept it as the explanation for the LF observations.

Finally we modeled numerically another possible scenario for the break
in the LF: could it be caused by the overlapping of \hii\ regions of a
certain size and above? In constructing the \hii\ rgion catalogue, one
of the difficulties that there is to overcome is that many \hii\ regions
appear to overlap on the image.  We adopted the solution proposed in
Rand (1992) of counting each peak in \ha\ as representing a single \hii\
region.  The flux of each \hii\ region was then estimated by integrating
over the pixels which could be reasonably attributed to a given region. 
One will undoubtedly miss a number of \hii\ regions that are too weak to
be detected in the vicinity of stronger emitters close by.  This will
influence the lower end of the LF but is not a significant factor in the
determination of the true LF at the higher luminosity end (Rand 1992). 
Anyway, to check directly if the overlapping cloud causes the break in
the LF, we applied a new model.  First we assumed the function in
eq.~(1) as the intrinsic LF.  To take care of the weaker regions in a
way corresponding closely to reality, we used eq.~(1) between 10$^{40}$
erg s$^{-1}$ and 10$^{37}$ erg s$^{-1}$ and below this limit we used a
flat distribution function ($\xi$ = 0 in eq.~(1)) following the observed
distribution in M31 by Walterbos \& Braun (1992).  In the computations
we allowed the luminosity of a region merged with others to reach an
upper limit of 10$^{42}$ erg s$^{-1}$, corresponding to the luminosity
of the regions formed by the mergers.  The distribution was discretized
into 300 bins of equal logarithmic width, and the volume of \hii\
regions per bin was calculated by simple summation, normalizing the
system so that an \hii\ region of radius 100 pc had an H$\alpha$
luminosity of log $L_{{\rm H}\alpha}$ = 38 (erg s\me) and calculating
the volumes in the first instance at infinite spacing between regions,
\ie\ with no overlaps.  The luminosity per bin was proportional to the
summed volume of regions in the bin, as we found observationally for the
regions below log $L_{{\rm H}\alpha}$ = 38 (erg s\me).  The total volume
of all the regions is $V_{\rm c}$.

In the model we put the regions whose integrated volume is $V_{\rm c}$
into a enclosure of volume $V$, so the overall filling factor is $V_{\rm
c}$/$V$.  The initial distribution is considered in finite luminosity
(\ie\ volume) bins, indexed as $i$, $j$, where $i$ and $j$ can both vary
from 1 to 300.  Interaction of the set of clouds, one in bin $i$ and the
other in bin $j$, is first considered. 

\begin{figure}[h]
\psfig{figure=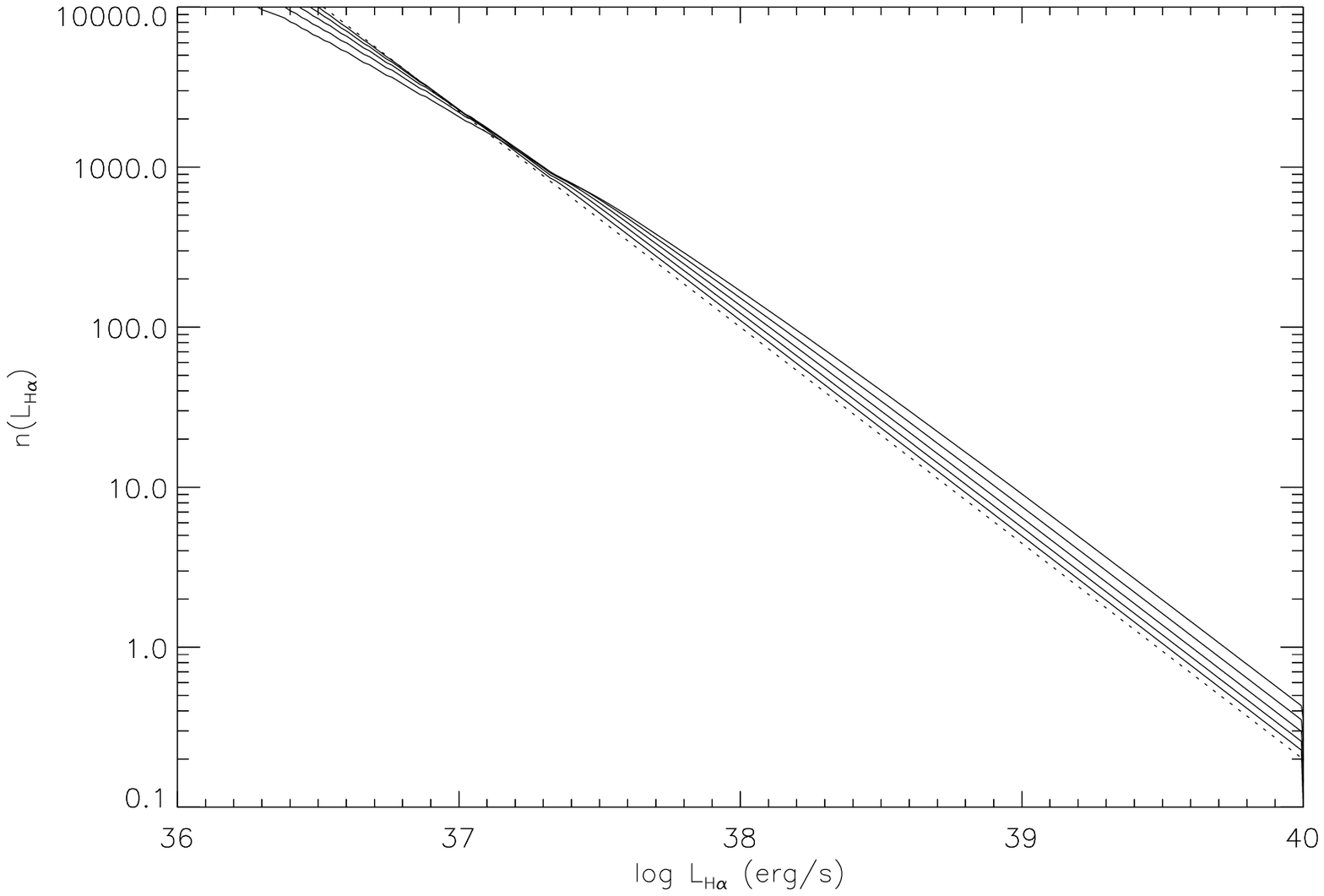,width=8cm}
\caption{Models of the LF in \ha\ in which increasing degrees of
overlapping and absorption take place between \hii\ regions (see text
for details).  Although we place little stress on the quantitative
results, qualitatively we can see that the effect of assimilation of
regions within others is to reduce the LF slope at high $L_{{\rm
H}\alpha}$, and not to increase the slope as observed.}
\end{figure}

The probability per unit volume, $P_j$, of a small cloud absorption
within a larger, is the filling factor of the clouds in bin $j$, which
is the same as using the rule that when two clouds are sited with the
distance between their centers less than the radius of the larger, the
smaller is absorbed into the larger, and the resulting \hii\ region has
the sum of the volumes (and luminosities) of the two.  The mean initial
spacing is determined by the chosen value of $V_{\rm c}$/$V$.  We obtain
for $P_j$:

\begin{equation}
P_j  =  \frac{N_j \, V_j}{V},
\end{equation}
where $N_j$ is the initial number of clouds within $j$, so the number  
$\delta N_f$ of clouds affected is just:

\begin{equation}
\delta N_f = N_i \, N_j \, V_j.
\end{equation}

We then adjust the luminosities in proportion to the revised 
numbers of clouds in each bin via:

\begin{equation}
\begin{array}{c}
N{_i}' = N_i - \delta N_f \\ 
N{_j}' = N_j - \delta N_f \\
N{_k}' = N_k + \delta N_f
\end{array}
\end{equation}

in which the total volume and luminosity are conserved, while the
regions affected by merging are transferred to the appropriate bin of
higher luminosity.  This procedure is carried out in order of increasing
bin luminosity, so that the merged regions participate in further
mergers if they are close enough to other regions.  In a single pass
combining all pairs of bins the new LF is computed.  modeling the
mergers in this way we assume that \hii\ regions are not optically
thick, so that only genuine three-dimensional overlaps, not projected
overlaps on the plane of the sky, need to be taken into account.  As
will be deducible considering Fig.~4, assuming optical thickness would
only accentuate the disagreement between this type of models and the
observed LFs.  In Fig.~4, it is clear that not only is there no evidence
of a glitch, but the negative slopes of the IMFs of the merged models
become less steep at high luminosities, in direct contrast to the
observations.  While it might be possible to find some type of
clustering model which would yield an LF similar to those observed, its
inputs would have to be very finely tuned, and would be unlikely to
reflect the realistic astrophysical conditions. 

In the light of the comprehensive theoretical study of \hii\ region LFs,
notably in the Galaxy, but with commentary on other galaxies, by McKee
\& Williams (1997, hereafter MW) we should add here that the effect of
density bounding on the LF in \ha\ which we claim here must function
independently of almost any detailed consideration of the IMF of the
stars which ionize the regions, or its temporal evolution.  The
truncation of the IMF at high stellar mass, adduced by MW to explain an
observed sharp steepening of the LF slope of \hii\ regions does not
conflict with the results presented here for two reasons.  Firstly the
phenomenon explained by MW occurs at an \ha\ luminosity almost an order
of magnitude below the glitch we find.  In the present study, this would
be very difficult to distinguish from the rather rapid change in slope
just above log $L_{{\rm H}\alpha}$= 37, due to the increasing
incompleteness of our statistics at low luminosity, but in any case it
is clear that one is dealing with two different real observable effects. 
Secondly a truncation of this type does not readily give rise to a
glitch in the LF, nor can it yield as a natural consequence the observed
sharp increase in internal brightness gradients and central surface
brightness with luminosity, described below in Section 2.2.  Thus
neither the observations nor the model of MW relate to the same
phenomenon as the one we describe here. 

\subsection{Central surface brightness and internal brightness gradient}

Further evidence for a change in physical regime at the \ha\ luminosity
$L_{\rm Str}$ comes from measurements of the surface brightnesses of
\hii\ regions.  The observations used are the same as those for the LFs:
photometrically calibrated CCD images of galaxies taken through an
appropriately redshifted \ha\ filter, with stellar continuum subtraction
via a neighboring off-band filter.  All observations were taken with the
4.2-m William Herschel Telescope on La Palma, and details of the
observations and the reduction procedure can be found in Rozas, Beckman,
\& Knapen (1996).  The absolute \ha\ fluxes and mean radii were found
for several hundred \hii\ regions per galaxy.  The largest, most
luminous, and isolated were selected for profile measurements, as these
subtend angles large enough to obtain reliable profile fits; a minimum
radius of 5 pixels was chosen.  This corresponds to an \ha\ luminosity
log $L_{{\rm H}\alpha}$ = 38 in the furthest galaxy observed. 

The numbers of regions per galaxy which could be used for these
measurements was 30--35.  Two characteristic parameters were derived per
region: the central surface brightness and the mean internal
surface-brightness gradient (this latter was also treated in Rozas,
Casta\~neda, \& Beckman 1998). 

\begin{figure*}[h]
\psfig{figure=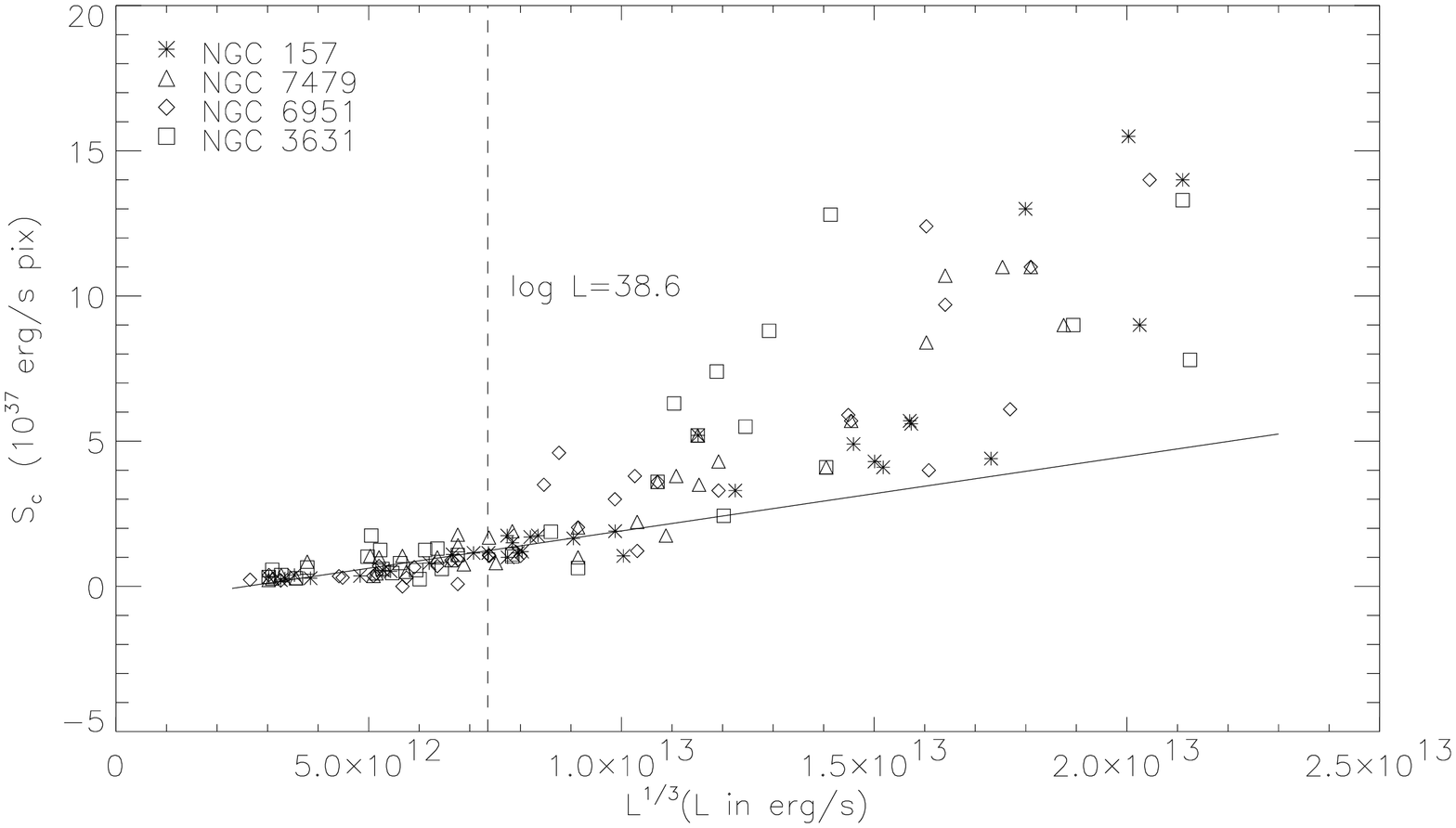,angle=90,width=15cm}
\vspace{2cm}
\psfig{figure=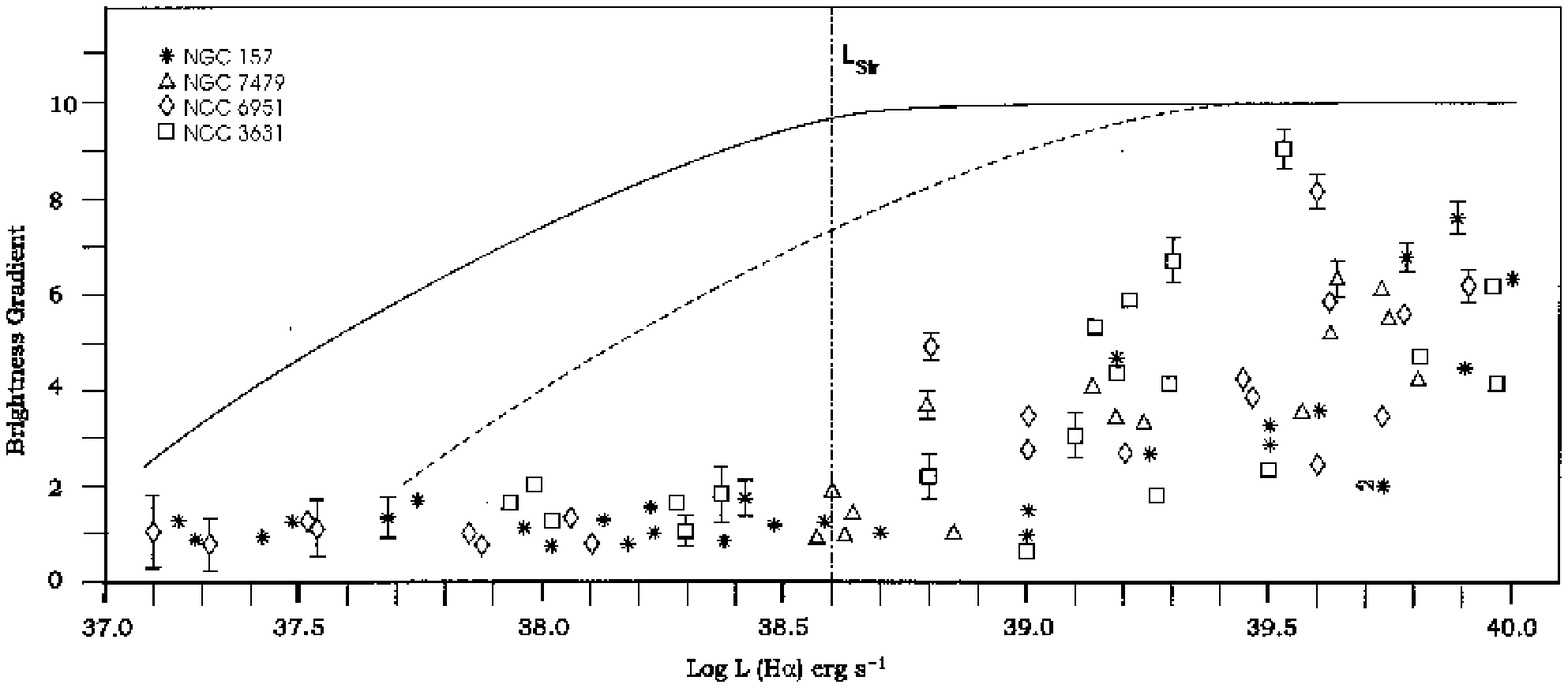,angle=270,width=15cm}
\caption{(a): Central surface brightness, $S_{\rm c}$, in H$\alpha$,
{\it vs.} the cube root of the H$\alpha$ luminosity, $L_{{\rm
H}\alpha}$$^{1/3}$, for \hii\ regions in four galaxies, selected for
minimum spatial blending with neighboring regions.  Below $L_{{\rm
H}\alpha}$$^{1/3}$ = 7.5 $\times$ 10$^{12}$ there is a good linear fit
to the data, predicted for ionization-bounded regions with a constant
product of density, electron density and filling factor.  Above this
value there is a clear change of regime; $S_{\rm c}$ rises much more
sharply with $L^{1/3}$; the scatter also widens considerably.  This
change must be caused by a change in the interactive behavior
gas--stars, and cannot be due only to changes in stellar content or
evolution within the regions concerned.  (b): Radial gradient parameter;
the ratio of the central surface brightness of a region to its estimated
radius, in linear units.  Near constant gradients are found for log
L$_{H\alpha}< $ 38.6, and strongly increasing values with increasing
scatter, for higher luminosities.  The curves show the ratios of
observed to intrinsic gradient for the two extremes of seeing-limited
resolution (0.8$''$ and 1.5$''$) during the observation.  A set of
regions with intrinsically constant gradients would show observed
gradients tending to a constant value at high $L_{{\rm H}\alpha}$,
rather than the steeply rising gradients observed.}
\end{figure*}

In Fig.~5a we show central surface brightness $S_{\rm c}$ as a function
of $L_{{\rm H}\alpha}^{1/3}$, the cube root of the \ha\ luminosity.  For
spherical ionization-bounded regimes with constant density and filling
factor, $S_{\rm c}$ should be proportional to the radius of a region,
and as luminosity will be proportional to volume, $S_{\rm c}$ should
vary linearly with $L_{{\rm H}\alpha}^{1/3}$.  We can see from Fig.~5a
that for regions with luminosities below $L_{{\rm H}\alpha}$ $\sim$
$L_{\rm Str}$ this relation holds rather well, but for luminosities
above this value there is a clear and sharp departure from
proportionality.  In Fig.~5b we show a brightness gradient parameter,
obtained by dividing the central surface brightness of a region by its
linear radius projected in the plane of the galaxy, plotted versus
$L_{{\rm H}\alpha}$.  We can see that below $L_{{\rm H}\alpha}$ =
$L_{\rm Str}$ this parameter varies very little, within the limits of
error, while above this luminosity it tends to rise.  Although the
scatter also increases, the trend to higher values is clear, and the
ratio of this gradient parameter above $L_{\rm Str}$ to the constant
value below $L_{\rm Str}$ reaches 10 in the steepest cases, and is
greater than 3 in most regions with log $L_{{\rm H}\alpha}$ $>$ 39.  The
change in behavior seen in Figs.~5a and 5b could not be due to the
finite resolution of the images.  It is easy to show that for a
hypothetical set of regions with constant intrinsic brightness gradient
the observed parameter, if degraded by instrumental resolution, would
tend asymptotically to a constant value at large radii, \ie\ at high
luminosity.  This is just the converse of what we measure: a tendency to
a constant value at low luminosity.  If on the other hand the intrinsic
gradient were in fact decreasing with increasing luminosity it would
take a ``conspiracy'' to flatten the plot at low $L_{{\rm H}\alpha}$
(\ie\ an inadequate deconvolution folded with an intrinsic fall would
need to cancel neatly, yielding a level plot), and the result would
disagree even more strongly with the observations at high $L_{{\rm
H}\alpha}$.  It would be very surprising to find the observed change of
gradient at a similar luminosity for all the galaxies sampled if it were
due to a resolution effect, since they are at quite different distances,
so the linear resolution on the galaxy disk corresponding to the angular
resolution of the telescope differs from object to object.  The observed
changes in the value of the gradient above $L_{{\rm H}\alpha}$ = $L_{\rm
Str}$ are way outside the most pessimistic error bars, so that Figs.~5a
and 5b offer prima facie evidence of a change in physical parameters
close to $L_{{\rm H}\alpha}$ = $L_{\rm Str}$.  The regions below this
luminosity obey a linear volume--luminosity relation, while those above
it in general do not.  This break in the geometrical properties of the
regions cannot be explained plausibly in terms of the collective
properties of the ionizing stars.  It is easy to show, and it was shown
by Str\"omgren himself, that for an \hii\ region forming within a medium
of constant uniform density (and, we should add here, constant filling
factor) the volume ionized is proportional to the rate of emission of
ionizing photons, so that the geometrical distribution of the brightness
within a region will not depend on the number of its ionizing stars,
provided that they are concentrated in a cluster near the center
(certainly true for all the regions studied in this section, which have
radii of well over 100 pc).  The observed changes must be due to the
behavior of the surrounding gas.  We will discuss in Sect.  3 of the
paper how these changes might occur for the observed regions.

A further point of interest here is that the constancy of the brightness
gradient parameter, and the dependence of the central surface brightness
on $L_{{\rm H}\alpha}^{1/3}$ below $L_{{\rm H}\alpha}$ = $L_{\rm Str}$
are not consistent with the dust model which we described in Sect.  2.1
as a testable alternative scenario for the observed behavior of the LF. 
Dust extinction increasing linearly with the radius of the \hii\ region,
and causing a notable bump in the LF close to log $L_{{\rm H}\alpha}$ =
38.6, would have an effective value of optical depth $\tau_{{\rm
H}\alpha}$ (at \ha) close to 0.6 at our normalization value of
log~$L_{{\rm H}\alpha}$ = 38, while $\tau_{{\rm H}\alpha}$ would be 0.4
at log~$L_{{\rm H}\alpha}$ = 37.5, and 0.9 at log $L_{{\rm H}\alpha}$ =
38.5.  The surface-brightness gradient would show an easily detectable
systematic fall between 37.5 and 38.5, by a factor of order 2, rather
than remaining constant as observed.  If above $L_{{\rm H}\alpha}$ =
$L_{\rm Str}$ the observed increase in the surface brightness were due
to dust being dispersed in these regions (perhaps by extra radiation
pressure), the gradient would not increase by factors of 3 to 10 from
its value at $L_{{\rm H}\alpha}$= $L_{\rm Str}$, but at most by a factor
2, and in any case it would never rise above the asymptotic value at
small $L_{{\rm H}\alpha}$, a rise which is dramatically observed.  Very
similar considerations hold for the central surface brightnesses.  We
will see in Sect.  4 that the diffuse \ha\ observed in the disks of our
galaxies offers more direct evidence that ionizing radiation is in fact
leaking out of the luminous \hii\ regions, and this tells against major
dust extinction in these regions.

\subsection{Internal velocity dispersion of the \hii\ regions}

The evidence discussed in the present sub-section does not refer to the
same set of galaxies as those treated in the bulk of the paper, because
the kind of data required, Fabry--Perot mapping of a galaxy in an
emission line, is in no way as easy to obtain as the narrow-band imaging
on which the majority of our deductions are based.  We have, to date,
analyzed Fabry--Perot maps in \ha\ of the disks of only two galaxies,
obtained in a program of disk-wide kinematics, not aimed principally at
the properties of \hii\ regions.  However the statistical results on the
internal velocity dispersions of the \hii\ regions observed are of such
relevance to the present discussion that it is important to include them
here.  They were made with the TAURUS instrument on the 4.2-m William
Herschel Telescope on La Palma.  Details of the observations and
reduction are presented elsewhere (Rozas, Sabalisck, \& Beckman 1998). 
We present the data here for two galaxies: M100 and M101.  We had
absolute flux calibrations for the regions in M100, obtained by our
group in a parallel context (Knapen 1998), but for M101 we have no such
absolute fluxes, nor are they available in the literature, so we give
the results in relative flux units.  This still allows us to infer the
essential relations between flux, $L_{{\rm H}\alpha}$, and internal
velocity dispersion, $\sigma$, for the regions of M101. 

\begin{figure*}[h]
\plottwo{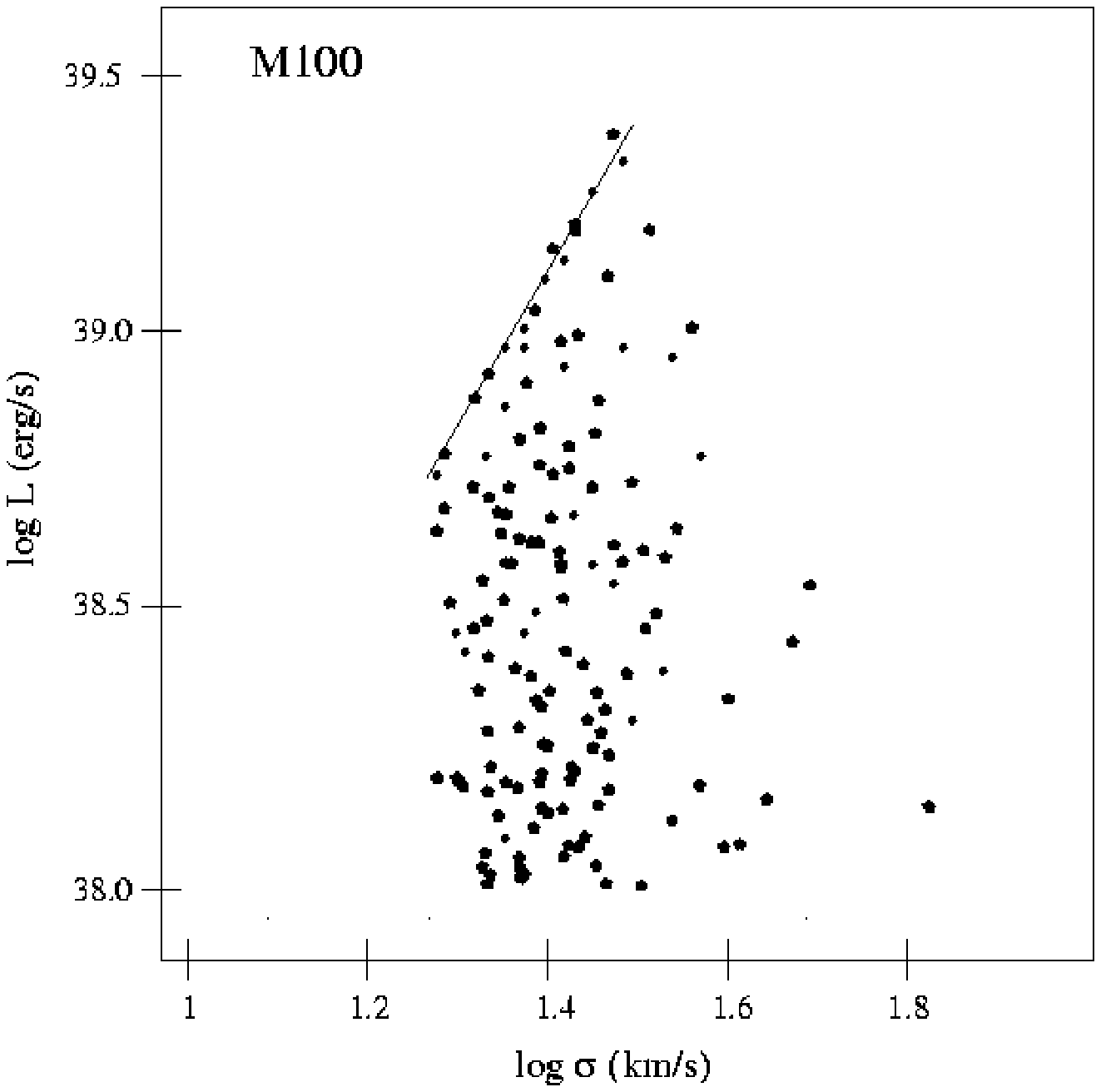}{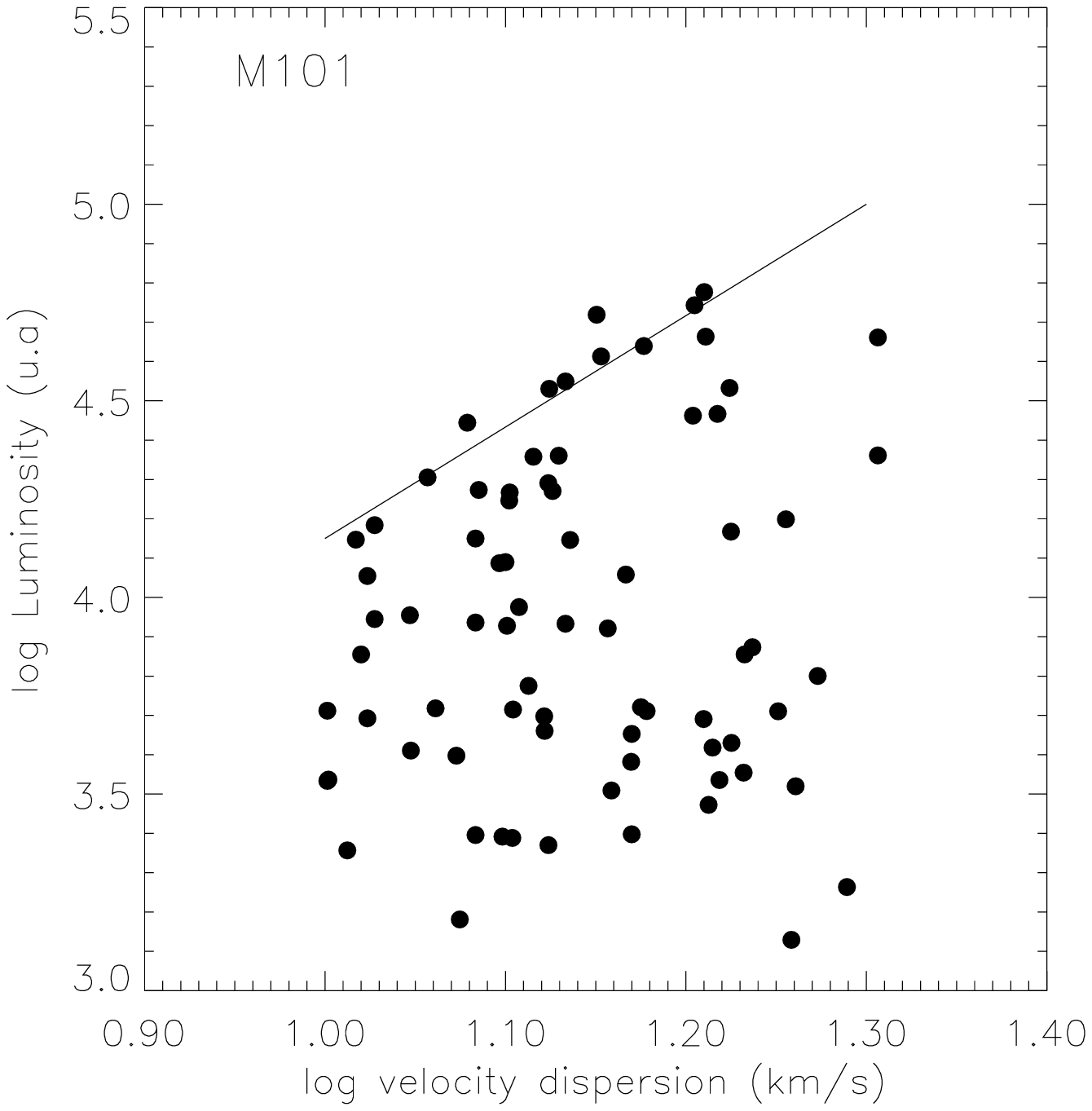}
\caption{(a): Log $L_{{\rm H}\alpha}$ {\it vs.\/} the velocity
half-width, $\sigma$, for the most intense emission components from the
\hii\ regions in M100, observed using the TAURUS Fabry--Perot
spectrometer.  The linear fit is to the upper envelope of the plot,
which should represent the locus of virialized regions.  The slope is
not 4, as predicted for ionization-bounded models, but 2.6, in agreement
with density-bounded models for regions with log $L_{{\rm H}\alpha}$ $>$
38.6.  The same slope was measured by Arsenault, Roy, $\&$ Boulesteix
(1990) using highest surface brightness rather than lowest $\sigma$ as a
selection criterion, in full accord with the hypothesis of density
bounding.  (b): Plot of log $L_{{\rm H}\alpha}$ {\it vs.\/} log $\sigma$
for the principal emission components in M101, from a TAURUS data cube
without absolute flux calibration.  A similar envelope at high $L_{{\rm
H}\alpha}$ and low $\sigma$ is found as for M100, with a slope
calculated at 2.55, in accord with the hypothesis that the most luminous
\hii\ regions in M101 are density bounded.}
\end{figure*}

Figure 6a is a plot of log~$L_{{\rm H}\alpha}$ {\it vs.} log~$\sigma$
for the principal components of the brightest \hii\ regions in M100. 
Only the brightest 10 $\%$, some 200 of the almost 2000 regions whose
absolute fluxes and diameters had been determined (Knapen 1998), were
bright enough to yield velocity profiles of high enough signal to noise
ratio to give well-measured values of $\sigma$ from single-frame
exposures.  We also have a lower limit cut-off in $\sigma$ of 10 km
s$^{-1}$, due to the combined uncertainties in the instrumental profile,
and the natural plus thermal broadening contributions.  At first glance
Fig.~6a is little more than a scatter diagram, and any attempt to put a
linear, or even a polynomial or spline fit through the points yields
little of value.  One might be tempted to conclude that the horizontal
spread is a little less than the vertical spread, so that a line of high
gradient, perhaps akin to the prediction based on the assumed virial
equilibrium of the \hii\ regions (Terlevich \& Melnick 1981),

\begin{equation}
\log \,{L}_{{\rm H}\alpha} = 4 \,  \log\, \ \sigma + c,
\end{equation}

could be brought to fit the points, but there is no real justification
for such a fit.  However there is one locus of regularity in these
points, namely the upper envelope of minimum $\sigma$, which can be well
fitted by a straight line of slope 2.6 ($\pm$0.14).  It is worth
comparing this with the linear fit to the log $L_{{\rm H}\alpha}$ {\it
vs.} log $\sigma$ plot for the \hii\ regions in M100 by Arsenault, Roy,
\& Boulesteix (1990) using a similar observational technique to ours. 
In their case they selected the regions of greatest surface brightness,
rather than an envelope of minimum $\sigma$.  This coincidence is very
significant physically, as we will show in Section 3.  In Fig.~6b we
show a similar log $L_{{\rm H}\alpha}$--log $\sigma$ plot for M101;
although the luminosity is not on an absolute scale, as explained above,
we can see here also the linear envelope to what would otherwise be a
scatter diagram.  The best fit measured slope to this is 2.55
($\pm$0.15), and the display of points in the graph is similar to that
in Fig~6a, with a slight trend for the $\sigma$ values to occupy a wider
range at lower luminosities. 

Our interpretation of the observations in Figs.~6a and 6b is that, in
general, the internal velocity dispersions of the \hii\ regions as shown
in their \ha\ emission line profiles are not in virial equilibrium. 
Whatever the mechanisms for transferring energy to the ionized gas (and
it is beyond the scope of the present article to consider these) it is
reasonable to assume that the emission line width, for an arbitrary
region, will be enhanced by the presence in the gas undamped effects of
the outflowing winds from massive stars, or even of supernova
explosions.  In the region NGC 604, in nearby M33, Sabalisck {\it et
al.\/} (1995) showed the morphological complexity of the velocity field. 
The principal velocity component for a more distant region, observed via
our technique, is a luminosity-weighted sum, which must incorporate any
discrete burst of input energy, and should not, in general, fit a virial
equilibrium relation.  In this scenario, regions showing the lowest
value of $\sigma$ for a given luminosity will be those whose emission
profiles are least affected by non-virialized energy injection, so that
regions closest to virial equilibrium should lie on, or near, the
low-$\sigma$ envelope of the log $L_{{\rm H}\alpha}$--log~$\sigma$
relation.  Physically, it is plausible that the most massive regions
will damp most rapidly the bursts of injected energy, so that in the
full log $L_{{\rm H}\alpha}$--log $\sigma$ diagram an increasing
fraction of regions will be found on the virial envelope with increasing
mass (and hence luminosity).  The linear envelopes in Figs.~6a and 6b
can thus be identified as the loci of those regions whose $L_{{\rm
H}\alpha}$--$\sigma$ relation should obey the conditions of the virial
theorem.

It is therefore very significant that the slopes of these envelopes are
$\sim$ 2.5 and not 4 as predicted for virialized systems (see, for
example, Terlevich \& Melnick 1981; Tenorio-Tagle, Mu\~noz-Tun\'on, \&
Cox 1993), since the value of 4 follows from assuming the conventional
picture of ionization-bounded regions.  The regions on the envelope in
Fig.~6a all have luminosities greater than $L_{\rm Str}$, so we would
expect them to be density bounded, and we will show below that a slope
close to 2.5 is in the range predicted for virialized regions which are
density bounded.  Arsenault, Roy, \& Boulesteix (1990) performed a
similar exercise to ours for the \hii\ regions of M100, but selected a
small number of regions of maximum surface brightness for their log
$L_{{\rm H}\alpha}$--log $\sigma$ plot.  The slope of their graph, 2.6,
confirms our scenario, as we have shown above that the regions of
highest surface brightness are found in the range $L_{{\rm H}\alpha}$
$>$ $L_{\rm Str}$, where the regions are most massive, where the virial
relation should hold in the ionized gas, and where our theory predicts
density bounding.

In this section we have presented evidence for a change in the physical
properties of the \hii\ regions in spirals in the very high luminosity
range (which were termed ``giant \hii\ regions'' by Kennicutt, Edgar, \&
Hodge 1989), above $L_{{\rm H}\alpha}$ = $L_{\rm Str}$ = 10$^{38.6}$ erg
s$^{-1}$ in \ha.  We have argued, based firmly on observation, that this
may mark a ``phase change'' from ionization bounding to density
bounding.  In the next section we argue from scaling calculations that
the occurrence of this type of change is physically plausible.

\section{Why should very luminous \hii\ regions be density bounded?}

\subsection{The luminosity function}

In Sect.  2.1 we outlined what would be required of the relation between
the mass in high-mass stars within a star-forming cloud and the mass of
the cloud itself so that the \hii\ regions with luminosities above a
specified value were density bounded, while those below this value were
ionization bounded.  The relation is that the mass in ionizing stars
should, with cloud mass, grow quickly enough for the ionizing flux to
rise more quickly than the cloud mass.  In the very schematic model we
present below we assume, to simplify the argument, that the average
cloud density does not vary, an assumption which holds well,
observationally, for regions up to the transition.  The model will not
fail physically if this condition is not well maintained to higher
luminosities, but the simplicity of the quantitative inferences will be
lost.

Let the ionizing flux, $L_{\rm i}$, from an \hii\ region depend on the
total stellar mass, $M_{\star}$, in ionizing stars within the region via

\begin{equation}
L_{\rm i} = k (M_{\star})^{\alpha},
\end{equation}

where $k$ is a constant.  We can set observational constraints on
$\alpha$.  From the semi-empirical study by Vacca, Garmany, \& Shull
(1996) we find that the rate of emission of Lyman continuum photons from
OB stars rises approximately as the square of the stellar mass.  The
index falls from values higher than 2 at early B and late O to values
close to 2 at O3.  For a uniform ``Salpeter" IMF slope, and no physical
cut-off in stellar mass at the upper mass end with increasing placental
cloud mass, the weighted mean index for a young star cluster would be a
little over 2, and this would be the appropriate value for $\alpha$. 
If, however, there were a fixed physical stellar mass limit, rather than
a statistical limit for each cluster, $\alpha$ would be unity or close
to unity.  Although evidence on this point is not abundant, Massey \&
Hunter (1998) measured with {\it HST\/} the photometric properties of
the most massive stars in the cluster R136 which ionizes the very
luminous \hii\ region 30 Doradus in the LMC.  They found, {\em inter
alia\/}, that the IMF slope appears to retain its Salpeter value of
--2.35 to the highest masses, and that there is no obvious evidence for
an upper mass cut-off; they find stellar masses ranging well over 120
M$_{\odot}$, through 130 M$_{\odot}$ and up to 150 M$_{\odot}$.  Without
generalizing unduly, or being especially optimistic, it is reasonable to
infer that $\alpha$, though not necessarily greater than 2, will
generally be higher than 1, and we will say provisionally that $\alpha$
should be between 1 and 2.

Now we let the mass $M_{\star}$ in stars contributing significantly to
the ionization depend on the placental cloud mass M$_{\rm Cl}$ according
to

\begin{equation}			
M_{\star} = j \,  M_{\rm Cl}^{\epsilon},
\end{equation}

where $j$ is a constant.  There is not very much direct evidence on the
value of $\epsilon$.  Larson (1982) reported, from a compilation of
observations within the Galaxy, that the highest mass of a young cluster
varies as the placental cloud mass to the power 0.43.  If we take the
IMF slope as --2.35, the mass of stars contributing to the ionizing
photon flux rises with an index of $\sim$ 0.43$\times$2.35; we derive
this value of 2.35 as a sum of 4.35 from the mass-weighted integral of
the Salpeter function, and --2 which takes into account the effective
contribution of mass to ionizing luminosity.  Using Larson's result thus
yields a value for epsilon of 1.01, effectively of unity.

The condition that the rate of production of ionizing photons fully
ionizes, and overflows the placental cloud of a young cluster for clouds
whose masses (and therefore \hii\ region luminosities) exceed a specific
value is found by combining (10) and (11), and is just

\begin{equation}			
\alpha  \epsilon > 1.
\end{equation}

The empirically inferred values for $\alpha$ and $\epsilon$ derived
above readily satisfy this condition.  Although, as we have seen, direct
evidence related to $\alpha$ and $\epsilon$ is sparse, there are two
indirect ways to estimate $\alpha \, \epsilon$ from the observations
presented in the present paper, both of which confirm that $\alpha \,
\epsilon$ must be greater than unity, one using the gradient of the \ha\
LF, and the other from the slope of the virial envelope of the log
$L_{{\rm H}\alpha}$--log $\sigma$ diagram. 

The first estimate comes from the ratio of the slopes of the \hii\
region LF in \ha\ above and below the critical luminosity $L_{{\rm
H}\alpha}$ = $L_{\rm Str}$. 

\begin{figure}[h]
\psfig{figure=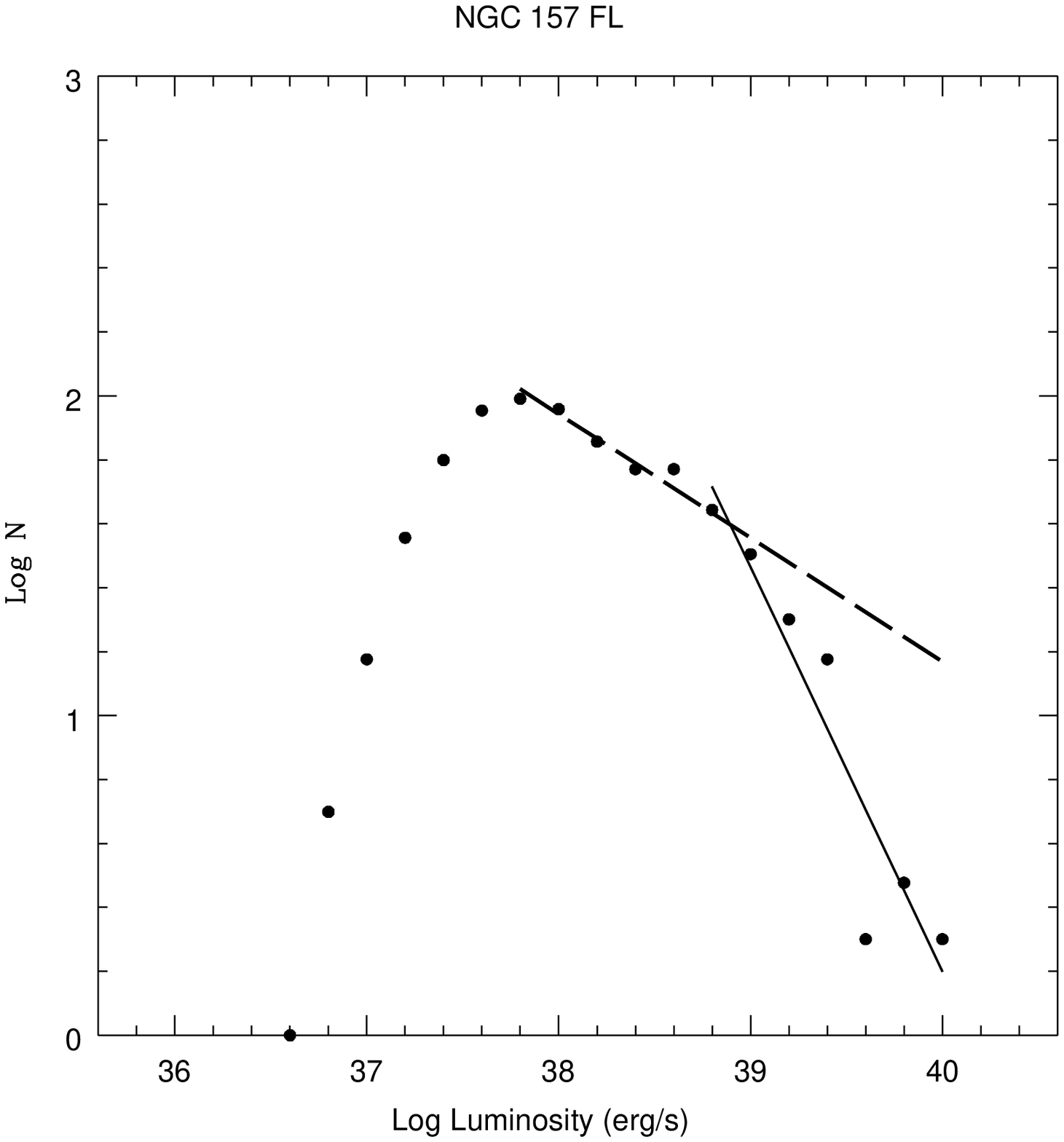,width=8cm}
\caption{Diagram showing how the integrated ionizing flux from the
luminous density bounded \hii\ regions in NGC 157 can be estimated by
subtracting the observed LF for log $L_{{\rm H}\alpha}$ $>$ 38.6 from
the extrapolated LF obtained by extending the observed function below
log $L_{{\rm H}\alpha}$ = 38.6 to higher values.  This estimate
indicates that the density-bounded regions leak sufficient Lyc photons,
in this galaxy, to ionize the diffuse medium without the need for other
sources of ionization.}
\end{figure}

We can express the LF for $L_{{\rm H}\alpha}$ $<$ $L_{\rm Str}$ as:

\begin{equation}
dN_{1}(L_{H\alpha}) = A \,  L_{{\rm H}\alpha}^{-\beta} \,  dL_{{\rm H}\alpha}
\end{equation}

and for $L_{{\rm H}\alpha}$ $>$ $L_{\rm Str}$ as:

\begin{equation}
dN_{2}(L_{{\rm H}\alpha}) = B \,  L_{{\rm H}\alpha}^{-\gamma} \,  
dL_{{\rm H}\alpha},
\end{equation}

where $A$ and $B$ are constants.  Inspection of Fig.~7 shows that this
parameterization is a good representation of the data.  In the range
$L_{{\rm H}\alpha}$ $<$ $L_{\rm Str}$ (below the transition), the
functional expression for the ionizing flux is the same that for the
\ha\ flux in equation (13) (as these \hii\ regions are ionization
bounded); so that from (10) we can write

\begin{equation}
dN_{1}(M_{\star}) = b \,  M_{\star}^{-\delta} dM_{\star},			
\end{equation}
where $b$ is a constant,  and 

\begin{equation}			
\delta = 1+\alpha(\beta-1).
\end{equation}

Combining (15) with (11) yields:

\begin{equation}
dN_{1}(M_{\rm Cl}) = c \,  M_{\rm Cl}^{-\eta} \,  dM_{\rm Cl},
\end{equation}
where $c$ is a constant,  and 

\begin{equation}
\eta = 1+\epsilon(\delta-1).
\end{equation}

Substituting (18) and (16) into (17) we have:

\begin{equation}			
dN_{1}(M_{\rm Cl}) = c \,  M_{\rm Cl}^{-(1+\epsilon \, 
\alpha(\beta-1))} dM_{\rm Cl}.
\end{equation}

Equation (19) is the mass function of the clouds below the transition 
($L_{{\rm H}\alpha}$ $<$ $L_{\rm Str}$).

If clouds with luminosities $L_{{\rm H}\alpha}$ $>$ $L_{\rm Str}$ are
fully ionized, the slope of their mass function is also that of their
luminosity function.  This slope is the index $\gamma$ of eq.~(14).  At
$L_{{\rm H}\alpha}$ = $L_{\rm Str}$ the slopes of the mass functions
below and above $L_{\rm Str}$ should be equal, as there is no reason to
predict a discontinuity in the cloud mass function.  Then we can write

\begin{equation}			
\gamma = 1+ \epsilon \,  \alpha(\beta-1),
\end{equation}
which is

\begin{equation}			
\alpha \,  \epsilon = (\gamma-1)/(\beta-1).	
\end{equation}

\begin{table*}[htb]
\begin{center}
\begin{tabular}{|l|c|c|c|}
\hline 
\hline 
Galaxy&$\log L_{\rm Str}$&$\beta$ ($L_{{\rm H}\alpha} < 
L_{\rm Str}$)&$\gamma$ ($L_{{\rm H}\alpha}>L_{\rm Str}$)\\
\hline
NGC 157$^{(1)}$ & 38.66 & 1.31& 2.18\\
NGC 3359$^{(2)}$& 38.60 & 1.44&2.10\\
NGC 3631$^{(1)}$ &38.54 & 1.51&2.16\\
NGC 5194$^{(3)}$ &38.63 & 1.52& 2.39\\
NGC 6764$^{(1)}$ &38.60 &1.30& 2.11\\
NGC 6814$^{(4)}$ &38.69 & 1.67 &3.39\\
NGC 6951$^{(1)}$ &38.49 & 1.33& 2.74\\
NGC 7479$^{(5)}$ (disk)&38.65 & 1.25& --2.10\\
\hline
\hline
\end{tabular}
\end{center}
\caption[]{The measured H$\alpha$ luminosities, $L_{\rm Str}$ erg
s$^{-1}$, at the peaks of the glitches, for the seven galaxies where
data of high quality yielding complete LFs in H$\alpha$ are available. 
Luminosities were measured using standard stellar calibrators, the
distance to the objects were taken on the basis of a value of 75 km
s$^{-1}$ Mpc$^{-1}$ for $H_{0}$, assuming that a measured radial
velocity is due exclusively to the Hubble flow In the succeding columns
are the slopes of the LFs in \ha\ below and above $L_{\rm Str}$,
$\beta$, and $\gamma$ respectively.  The mean measured uncertainty in
$\beta$ is $\pm 0.05$ and in $\gamma$, $\pm 0.15$. 
\small{$^{(1)}$ Rozas, Beckman \& Knapen 1996;} 
\small{$^{(2)}$ Rozas, Zurita \& Beckman  2000;} 
\small{$^{(3)}$ Rand 1992;} 
\small{ $^{(4)}$ Knapen {\it et al.\/} 1993;}
\small{$^{(5)}$ Rozas {\it et al.\/} 1999}}
\end{table*}

From eq.~(21) we deduce that the condition $\alpha \, \epsilon > 1$
implies $\gamma > \beta$, \ie\ that the slope of the LF in the range
$L_{{\rm H}\alpha}$ $>$ $L_{\rm Str}$ should be steeper than the slope
in the range $L_{{\rm H}\alpha}$ $<$ $L_{\rm Str}$.  We can see in
Fig.~1 that the LFs for all the galaxies observed do meet this
condition, \ie\ for each galaxy $\gamma$ is greater than $\beta$.  In
Table~1 we give numerical values for $\beta$ and $\gamma$ for eight
galaxies taken from our own work and from the literature.  Since
$\epsilon$ $\sim$ 1, we could use eq.~(21), and the measured LF in \ha\
for a galaxy to obtain an empirical value for $\alpha$.  Another way to
derive $\alpha$ is by weighting the data in Vacca, Garmany, \& Shull
(1996) by the stellar IMF.  Equating the two values enables us to derive
the IMF slope.  Using this technique, the values found lie in the range
2 to 2.5 for the LFs of the galaxies measured in the present paper. 
However the method is not intrinsically very accurate, as it requires
the ratio of two observables, each of form ($x-1$) where $x$ $\gg$ 1. 
Another, relatively more reliable way to obtain the IMF slope uses the
gradient of the log $L_{{\rm H}\alpha}$--log $\sigma$ envelope in the
velocity-dispersion diagram, as we will now explain in Section 3.2

\subsection{ The internal velocity} 

The virialization of the gas motions within \hii\ regions has been
proposed (Terlevich \& Melnick 1981) not so much as a working
hypothesis, rather as a desirable case, which would allow us to measure
the gravitational mass of an emitting region.  We showed in Sect.  2.3
that the majority of the \hii\ regions in the two galaxies whose log
$L_{{\rm H}\alpha}$--log $\sigma$ diagrams are plotted are not in virial
equilibrium.  The \ha\ flux they emit is a combination of individual
sources within each region, some of which have non-random motions with
relatively high velocities.  Although the mass fraction in these,
typically expansive, motions may not be high, a luminosity-weighted
profile incorporates them, and will always tend to have a super-virial
value for its half-width.  We argued in Sect.  2.3 that the lower
envelope in $\sigma$ to the log $L_{{\rm H}\alpha}$--log $\sigma$
distribution should correspond to the virialized regions, and showed
that the regions found on this lower bound in $\sigma$ (which is also an
upper bound in $L_{{\rm H}\alpha}$), will in general be density bounded. 
One corollary of this is that the predicted 4$^{\rm th}$ power
relationship between luminosity and $\sigma$, though good for the Lyc
flux of the regions on the envelope, should not hold for \ha\, since
\ha\ does not measure the total emitted flux in this case.  A
first-order estimate of the appropriate index relating both the total
\ha\ emitted flux and the Lyc flux can be obtained from the scaling
model in Section 3.1.  If the \ha\ luminosity of an \hii\ region above
the Str\"{o}mgren transition is $L_{{\rm H}\alpha}$, the Lyc continuum
luminosity (in units of the transition luminosity values) is:

\begin{equation}
L_{\rm Lyc}	= {L_{{\rm H}\alpha}}^{\alpha \,  \epsilon}
\end{equation}
and incorporating this relation into the virial prediction,  eq.~(9),  we
have 

\begin{equation}
 L_{{\rm H}\alpha}  = p \ \sigma^{4/\epsilon \, \alpha}  = p \ \sigma^{n},
\end{equation}

where $p$ is a constant, and $n = 4/ \epsilon \, \alpha$.  We can see
that the condition for density bounding, $\epsilon \, \alpha > 1$
implies that $n < 4$, and that the observed values for M100 and M101: $n
\sim 2.5$ are in agreement with this condition, yielding a value of
$\alpha \, \epsilon$ close to 1.6.  Equation (23) gives, in principle, a
more reliable method for measuring $\alpha \, \epsilon$ than the LF
slope ratio in eq.~(21), but it requires a much higher investment in
observation and reduction.  If we can combine the two tests for a number
of galaxies, this will give a useful quantitative check on the
density-bounding hypothesis.

\subsection{The stability of the transition luminosity}

In Table 1 we list the luminosities $L_{\rm Str}$ of the peaks in the
observed LF at the transition.  The mean value of $L_{\rm Str}$ is 4.05
$\times$ 10$^{38}$ erg s$^{-1}$, which is 1.4 $\times$ 10$^{50}$ photon
s$^{-1}$.This is the equivalent of ten main sequence stars of type O7,
or 2 of type O3 (Vacca, Garmany, \& Shull 1996), which should be raised
by a factor of 2.23 to allow for the Lyc/\ha\ ratio.  The rms scatter is
surprisingly low: 6.2 $\%$ of $L_{\rm Str}$, \ie\ 0.07 stellar
magnitudes.  It is important to note here that if the peaks were induced
by some effect of limited angular resolution, the rms scatter would be
0.4 magnitudes, due to the range of distances of the galaxies observed,
a powerful reason to reject this possibility, recently proposed by
Pleuss \& Heller (2000).  It is interesting to compare this with the
dispersion of the integrated luminosities of the same galaxies, taken
from the Tully catalog (1987), which show a scatter of 0.15 mag. 
Although the sample is too small to draw a powerful conclusion, the
promise of the glitch as a distance indicator is supported by the fact
that its scatter is significantly lower than that of the absolute
luminosities of the host galaxies in the sample under observation.

\placetable{Table 1}

Our sample is still too small to determine whether the transition can be
used as a standard candle on extragalactic scales, but it appears {\it a
priori} to satisfy the four criteria proposed by Aaronson \& Mould
(1986) for a cosmological standard candle.  It should: (a) exhibit
small, quantifiable dispersion; (b) be measurable in enough galaxies to
be calibrated locally; (c) have a well-defined physical basis, and (d)
be bright enough to be used well into the region where the Hubble flow
predominates.  The observed peak in the \ha\ LF for \hii\ regions in
disk galaxies promises to satisfy all four criteria.  We have given here
the empirical case for (a), (b), and (d), and very reasonable backing
for (c).  A criterion not brought out by Aaronson and Mould is that the
observable should be permanently available to observe.  This is of
course not satisfied by supernovae, and although permanency is not an
absolute requirement, the Str\"{o}mgren transition appears to be a high
luminosity and also a permanent feature of disk (and probably irregular)
galaxies.  The initial work presented here clearly requires support from
a larger sample of galaxies.

Although a purely empirically determined value of $L_{\rm Str}$ could,
given its low scatter, allow it to be used as a standard candle, it is
clearly desirable to establish the underlying physics more firmly than
we have been able to do here.  Nevertheless, if we can assume the
transition hypothesis maintained in the present paper it does give a
clear physical explanation for the low scatter in $L_{\rm Str}$. 
Starting from eq.~(10), and assuming a constant cloud density (which, as
explained above in Sect.  2.2 appears valid for clouds up to and
including those with $L_{{\rm H}\alpha}$ = $L_{\rm Str}$) we can rewrite
the equation in terms of the mass, $M_{\rm i}$, of the ionized cloud:

\begin{equation}			
M_{\rm i} = k' \,   M_{\star}^{\alpha},
\end{equation}

where $k'$ is a constant.  The stellar mass, $M_{\star}$, and the mass,
$M_{\rm Cl}$, of the whole placental cloud are related by equation (11). 
At the transition, the mass of the ionized cloud just equals the whole
cloud mass, and calling this mass $M_{\rm Str}$ we can find an
expression for $M_{\rm Str}$ by combining (24) and (11):

\begin{equation}
M_{\rm Str} =  k'^{(1/(1-\alpha \,  \epsilon))} \times 
j^{(\alpha/(1-\alpha \epsilon))},
\end{equation}
where $j$ is the constant of equation (11).

If we take $\alpha$ = 1.5 = $\alpha \, \epsilon$, the fractional variation in 
the mass of a cloud for a given luminosity will be:

\begin{equation}
\left|\frac{dM_{\rm Str}}{M_{\rm Str}}\right| = \left|\frac{-1}{2} 
\frac{dk'}{k'} 
\right|+ \left|\frac{-3}{4} \frac{dj}{j}\right|.
\end{equation}

However, in any given galaxy, the transition peak in the luminosity
function is formed by a significant number of regions, which we will
specify as $N$, so that the variance in the peak luminosity will be
given by:

\begin{equation}
\frac{dL_{\rm Str}}{L_{\rm Str}} = N^{-1/2} \,  
\frac{dM_{\rm Str}}{M_{\rm Str}} = N^{-1/2} \,  
\left( 0.5 \frac{dk'}{k'} + 0.75 \frac{dj}{j}\right) .
\end{equation}

Since in the galaxies observed $N$ is of order 50, $N^{-1/2}$ is of
order 0.15, and it is this collective property of $L_{\rm Str}$ which
makes it such a stable index as opposed to say the luminosity of the
$n^{\rm th}$ brightest individual \hii\ region.  We would expect the
main variable influencing both $k'$ and $j$ to be the metallicity.  The
parameter $j$ depends on the IMF, and recent results show, to the
surprise of some, that the metallicity apparently hardly plays a role in
determining the cluster IMF slope.  The IMFs of stars in the Galaxy
(Massey, Johnson, \& de Gioia-Eastwood 1995) and in the LMC (Massey {\it
et al.\/} 1995), with masses greater than 10 M$_{\odot}$ are found to
have slopes close to the Salpeter (1955) value of --2.35, in spite of
their large metallicity difference.  In a recent review article,
Elmegreen (1997) argues that this is because of the essentially fractal
nature of the fractionation process leading to star formation.  Whatever
the physical cause, this invariant property works in favor of an
invariant luminosity for the Str\"{o}mgren transition. 

The parameter $k'$ relating the number of Lyc photons to stellar mass is
of course a function of metallicity, but over a range between 0.5 solar
and 2 solar, calculations by Garc\'\i a Vargas et al.  (1995) show that
the flux from an O star will not vary by more than a few per cent, and
the quantitative effect of different metallicities on the net absorption
within the \hii\ regions themselves are also small, if we are looking at
hydrogen ionization (the situation would be more metal dependent for
helium).

In practice, to establish $L_{\rm Str}$ as a standard candle requires
further steps: calibration locally via Cepheid distances using the {\it
HST\/} key project galaxies and extension of the LF data base in nearby
galaxies to obtain a reliable variance for $L_{\rm Str}$.  The potential
range in $z$ would be limited by the angular sizes of the \hii\ regions,
rather than by their luminosities.  A practical ground-based limit is
given by the distance of an \hii\ region with $L_{{\rm H}\alpha}$ =
$L_{\rm Str}$ which subtends 0.5 arcsec, \ie\ 2$\times$10$^4$ km
s$^{-1}$.  The use of {\it HST\/} would push this limit back
considerably.  Since the variance appears to be low, this extra
secondary distance standard may be of use, \eg\ in probing the
depth--structure of neighboring galaxy clusters.

\section{Escaping Lyc photons, and the diffuse \ha\ emission}

The origin of the photons which ionize the diffuse interstellar medium
relatively far from \hii\ regions, which does not contain observable
sources of ionization has not been definitively assigned.  The OB stars
in \hii\ regions are in some sense a natural source of this ionizing
radiation, but doubts have arisen for two reasons.  Firstly can enough
radiation escape from the immediate environments of the OB stars, \ie\
from the \hii\ regions themselves, and secondly can the mean free path
of these photons be sufficiently long for them to ionize the diffuse
medium over the full extent of a galactic disk? These doubts have led to
at least one alternative model, notably than by Sciama (1990) who
proposed the decay of massive neutrinos as a means of producing {\it in
situ} Lyc photons in the diffuse ISM.  The hypothesis of density
bounding for the most luminous regions can resolve the first of the two
doubts (the second would have to await detailed models of the
inhomogeneous ISM for its resolution).  We can estimate the flux leaking
out of the luminous \hii\ region population by extrapolating the \ha\ LF
of the \hii\ regions in a galaxy in the range $L_{{\rm
H}\alpha}$~$<$~$L_{\rm Str}$ to the range $L_{{\rm H}\alpha}$ $>$
$L_{\rm Str}$, and subtracting off the measured LF in this upper range. 
The difference should approximate the ionizing flux equivalent which
escapes from the density-bounded regions.  If this is greater than the
measured diffuse flux, we have a {\it prima facie\/} case in favor of
the model as an explanation for the latter. 

\begin{figure}[h]
\psfig{figure=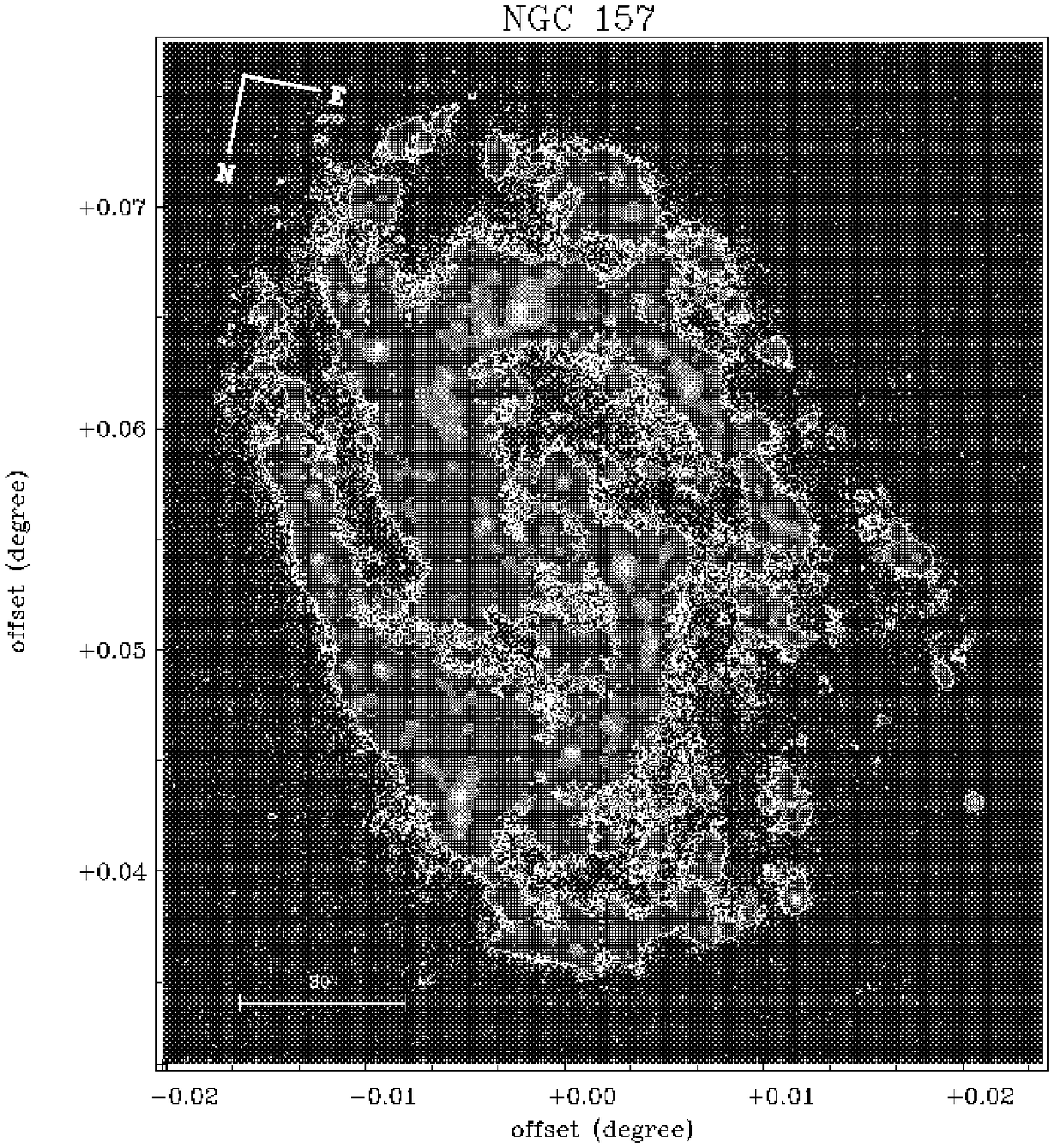,width=8cm}
\caption{H$\alpha$ image of NGC 157 illustrating how the regions of
highest luminosity also dominate in surface brightness, and how the
diffuse H$\alpha$ in the disk is correlated spatially with these.  The
major fraction of the ionization of the diffuse warm medium in disks
should be due to these extremely luminous, and also density-bounded
regions.  The external isophote is at an emission measure value of 7 pc
cm$^{-6}$.}
\end{figure}

We have used NGC~157 as an example to show whether this hypothesis is
worth pursuing, and Figs.~7 and 8 illustrate the procedure for this
galaxy.  To take into account \hii\ regions below our detectability
limit, we extrapolated our LF below this limit, using the LF measured
for M31 by Walterbos \& Braun (1992), which goes down to log $L_{{\rm
H}\alpha}$ = 35 (erg s$^{-1}$) thanks to the proximity of M31.  We
scaled the measured LF for NGC~157 at log $L_{{\rm H}\alpha}$ = 38 where
our sample is statistically complete, and used the form of the LF from
Walterbos \& Braun to extrapolate to lower values.  The comparison of
the observed truly diffuse \ha\ flux with that available due to Lyc
photons escaping from the density-bounded \hii\ regions is:

\begin{itemize}

\item Observed upper bound to the diffuse flux: 2.3 $\times$ 10$^{41}$
erg s$^{-1}$ (implied ionizing photon flux 1.7 $\times$ 10$^{53}$ Lyc
photon s$^{-1}$). 

\item Computed escaping flux: 3.1 $\times$ 10$^{41}$ erg s$^{-1}$ (2.3
$\times$ 10$^{53}$ photon s$^{-1}$).

\end{itemize}
	
This example shows that, on the density-bounding hypothesis, the flux
escaping from the luminous \hii\ regions is easily sufficient to yield
the diffuse \ha\ observed in NGC~157, and in a related study (Zurita,
Rozas, \& Beckman 2000) we are amplifying the sample treated in this way
to see whether the model is as good at accounting for the diffuse \ha\
in a wider sample of galaxies.  Ferguson {et al.\/} (1996) showed that
there is a good geometrical correlation between the positions of \hii\
regions in general and the diffuse flux, in the two galaxies they
studied in detail.  It will be interesting to see if this correlation is
improved when the most luminous \hii\ regions are used, as should be the
case if our model is valid.  At this stage, given the clumpiness of all
\hii\ regions, we would not postulate that Lyc photons cannot escape at
all from regions in the range $L_{{\rm H}\alpha}$ $<$ $L_{\rm Str}$,
only that the regions above $L_{\rm Str}$ should be contributing
proportionally in excess of their already high measured \ha\ output.

It is interesting that the computed escaping flux in NGC~157 is in
excess of the observed diffuse flux.  This difference appears to be well
outside our estimates of random and systematic errors in all the
quantities estimated, and suggests that the down-conversion of Lyc
photons in the diffuse medium is by no means sufficient to soak up all
the ionizing flux leaking out of the density-bounded regions.  A good
fraction of all the ionizing photons emitted by the OB stars must escape
completely from the plane of the galaxy.  If this effect is true of the
majority of disk galaxies, the implication for the ionization of the
intracluster medium in clusters of galaxies is of considerable potential
interest.

\section{Discussion}

We have used three types of observations to argue for a change in
physical regime between ionization bounding and density bounding in
\hii\ regions at a characteristic \ha\ luminosity close to log $L_{{\rm
H}\alpha}$ = 38.6 erg s\me\, which we have termed the Str\"omgren
luminosity, $L_{\rm Str}$.  They are the change of slope, accompanied by
a ``glitch'' in the \ha\ LFs of the complete \hii\ region populations in
a set of observed disk galaxies, the quantitative relation between
internal velocity dispersion, $\sigma$, and \ha\ luminosity, $L_{{\rm
H}\alpha}$, of the \hii\ regions on the virial envelope of the log
$L_{{\rm H}\alpha}$--log $\sigma$ distribution for the complete
population in a galaxy, and the change in behavior of the \ha\ surface
brightness of regions in the range $L_{{\rm H}\alpha}$ $\approx$ $L_{\rm
Str}$.  Of these, the first two are presented as evidence of the kind of
transition proposed, while the third, though of interest, is not as
precisely defined by the observations, and its cause is at this stage
fully open to other explanations.  Indeed in order to sustain our basic
hypothesis we will clearly require spectroscopic observations, as
specified below.  However, the narrow range in the luminosity of the
peaks of the LFs presented here means that this feature is of
considerable interest on purely empirical grounds, as a potential
secondary standard candle.  For this reason alone it is worth making a
firm effort to understand the underlying physical behavior causing the
peak and change of slope. 

In Sect.  2.1 we showed that two phenomena which could occur, and which
at first thought might be able to account for the LF observations (the
effect of the overlapping of regions on scales commensurate with the
luminosities around $L_{{\rm H}\alpha}$ = $L_{\rm Str}$, and the effect
of dust extinction within and enveloping the regions) will not give rise
to the observed LF change, and are not in fact viable as explanations
for the observations.  We ought also to consider an effect brought out
by MW, who examine the relationship between the properties of \hii\
regions within the Galaxy, and their OB stellar components.  They
present models to explain the change in slope of the \ha\ LF at
moderately high luminosities, which they attribute to an essentially
statistical effect due to the discrete number of high-luminosity stars,
and its relation with the total ionizing luminosity of an \hii\ region. 
In their paper, MW discuss the effects of envelopes of diffuse gas
around giant \hii\ regions, which are observed for the majority of
regions, and which correspond to the ``core--halo'' structure noted in
Kennicutt, Edgar, \& Hodge (1989).  In the present study we have taken
this structure as correct, and have not discussed it in further detail. 
The regions defined by our empirical limiting isophotal method contain
the whole of the core and the halo, and are clearly bigger than regions
defined only via their cores, or via their emission radii at radio
wavelengths.  It is important to note here that the change in the LF
slope predicted by the MW models occurs at an \ha\ luminosity an order
of magnitude less than that of the glitch we observe.  Our observations
here effectively disguise the MW break, because it occurs where our
sample is incomplete, leading to the broad peak and decline to low
luminosities which are statistical artifacts.  Neither our models nor
our observations contradict the results of MW, but their work does not
examine the LF gradient change at $L_{{\rm H}\alpha}$ = $L_{\rm Str}$. 

The changes which occur in the surface brightness and surface-brightness
gradients of the regions in the luminosity range $L\approx L_{\rm Str}$
are clearly of interest physically, but it is not so easy for us to show
that they are attributable to a transition to density bounding, although
we do have significant technical advantages in resolution and signal to
noise over the observers in the 1980s (for example the number of regions
listed in Kennicutt, Edgar, \& Hodge 1989 for NGC~7479 was 67, while the
in data yielding the internal brightness gradients used in the present
paper we catalogued over 1000 regions in this galaxy (Rozas {\it et al.}
1999).  Nor can the changes in parameters presented in Figs.~5a and 5b
be readily attributed merely to angular resolution limitations, as
suggested recently by Pleuss \& Heller (2000), based on their study of
M101.  Their argument is based on a scaling of the results from that
galaxy to a distance of 20 Mpc, but the distances of the galaxies in the
present sample range from 7 to 38 Mpc ($H_{\rm 0}$ = 65 km s\me\
Mpc\me), a factor of over 5 in distance.  The observed increase in
surface brightness and surface-brightness gradients in the range
$L_{{\rm H}\alpha}$ $>$ $L_{\rm Str}$ cannot occur for regions with
constant density and filling factor, as well pointed out by McCall
(1999); density-bounded regions would not show surface-brightness
gradients different from ionization-bounded regions if their mean
densities and filling factors were invariant.  Thus although we have
used the approximation of constant density and filling factor in
modeling the changes in properties across the transition, it cannot hold
for regions at much higher luminosities than $L_{\rm Str}$.  The product
of density and filling factor must rise in this range.  We can examine a
simple case in which two \hii\ regions have the same \ha\ luminosity but
one is ionization-bounded and the other density bounded; the factor
relating the total ionizing luminosity of the density-bounded region to
the fraction which is down-converted to \ha\ we term $g$.  In a
self-consistent scenario, $g$ can represent the factor by which the
product of the density and filling factor of the density-bounded region
exceeds that in the ionization-bounded region.  It is easy to show that
the central surface brightness of the (assumed spherical)
density-bounded region will be the greater by a factor $g^{2/3}$, and
the radial-brightness gradient, parametrized as the central brightness
divided by the radius, will be greater by a factor $g$.  This exercise
does not prove that density-bounded regions will necessarily have higher
surface brightnesses, and brightness gradients, only that the assumption
of density bounding is consistent with this condition, and hence
consistent with our observations.  In order to fulfill this requirement,
either the density, or the filling factor, or both, must be rising
parameters above $L_{{\rm H}\alpha}$ = $L_{\rm Str}$.  While it is
perfectly plausible that more intense ionizing sources produce higher
filling factors, and that the most massive clouds have higher mean
densities, it is not obvious that these changes will not begin to occur
below $L_{{\rm H}\alpha}$ = $L_{\rm Str}$, \ie\ it is not obviously a
physical necessity that the surface-brightness increase occur just at
the transition.  The fact that it appears to occur at, or near, what we
believe to be the transition cannot at this stage be taken as strong
support for the hypothesis of density bounding.  Direct, spatially
resolved measurements of in situ electron densities via line intensity
ratios will be required to further our understanding of this point.  We
must also be prepared to examine our data in the light of non-isotropic
models for luminous \hii\ regions such as the ``chimney'' hypothesis
(Norman \& Ikeuchi 1989; Heiles 1990) discussed in more physical detail
by Dove \& Shull (1994), in which at a critical luminosity a region can
break out of the denser disk gas into the halo.  These models appear,
however, to predict reduced surface-brightness gradients for regions
observed in face-on galaxies, and so are unlikely to offer an
explanation of our observations. 

One of the most attractive aspects of our hypothesis is that it offers a
promising scenario to account for the diffuse \ha\ from gas dispersed
over the disk of a galaxy, outside the \hii\ regions.  Two quite recent
studies have pointed to the emission of ionizing flux from OB stars as a
satisfactory explanation of the \ha\ emission from the warm ionized
medium in disk galaxies.  Oey \& Kennicutt (1997) made a direct
comparison of the rate of emission of ionizing photons from the OB stars
in the Large Magellanic Cloud, with the rate of \ha\ emission from \hii\
regions, and concluded that up to 50 \% of the flux emitted by the stars
is not down-converted within the \hii\ regions, and is thus available in
principle to ionized the warm diffuse medium.  They compare this with an
estimate of 35 \% of the radiation emitted by the diffuse medium in
\ha\, and conclude that the OB stars do put out sufficient flux in this
galaxy.  A study by Ferguson et al.  (1996) of NGC~247 and NGC~7793
concluded that the integrated flux required to ionize the diffuse gas is
some 40 \% of the total \ha\ flux emitted by a galaxy, including its
\hii\ regions.  However, this calculation assumed, as have any such
previous calculations, that the total flux of ionizing photons from an
\hii\ region can be directly measured via its \ha\ flux.  If the
hypothesis presented in the present paper is correct, a significant
fraction of the total ionizing flux liberated within the \hii\ region
population of a galaxy is not converted to Balmer radiation within the
\hii\ regions.  This is true of the most luminous regions, with
luminosities ranging up to 10$^{40}$ erg s\me, and the escaping flux can
range up to rather more than 10$^{40}$ erg s\me\ for the brightest
regions.  The integrated escaping flux for a galaxy can attain a few
times 10$^{41}$ erg s\me, easily enough to account for the measured
diffuse \ha\ flux, according to the observations of Ferguson et al. 
(1996) and of our own (Zurita, Rozas, \& Beckman 2000).  Ferguson et al. 
(1996) pointed out that the emitting diffuse gas tends to surround \hii\
regions, and we can go further on this point, noting the geometrical
correlation with the most luminous regions, which we will go further to
quantify in a new study (Zurita, Rozas, \& Beckman 2000).  Finally if
our hypothesis is valid, a significant fraction of all the ionizing
photons emitted by the OB stars in a galaxy may escape completely from
the disk into the halo and finally into the intergalactic medium.  Two
corollaries are of interest here: intergalactic clouds, even relatively
far from a major galaxy, may have their surface layers ionized by this
escaping flux, and global star formation rates in disk galaxies,
estimated from integrated \ha\ fluxes, may in fact be significantly
larger than these estimates.  Both of these points are worth pursuing
theoretically and observationally. 
	
We should treat the density bounding-hypothesis for the luminous \hii\
regions as speculative.  Fundamentally missing pieces of the puzzle are
line-ratio tests, of the sort performed by McCall, Rybski, \& Shields
(1985) comparing [\oiii] to [\oii] emission-line intensities (in a
density-bounded region, the [\oii] Str\"omgren sphere should under many
conditions be larger in radius than the \hii\ region, so the
[\oiii]/[\oii] ratio should be significantly enhanced over its value in
ionization-bounded regions.  McCall, Rybski, \& Shields (1985) claimed
one such detection, region(-606-1708) in NGC~598, but no others, and
this detection has since been called into question (McCall 1999). 
Technical advances of the type represented by the TTF (Taurus Tuneable
Filter, Bland-Hawthorn \& Jones 1998) which permits imaging of complete
galaxies in single emission lines with full redshift flexibility, will
allow us to apply line-ratio tests over complete populations of fully
imaged \hii\ regions in galaxies such as the ones measured only in \ha\
for the present paper.  Another prospect, but only for the nearest
galaxies due to limitations on angular resolution, is the use of
two-dimensional fiber-fed spectrographs to sample the emission across
the full face of an \hii\ region.  Here we would trade complete spectral
coverage for limitation on the number of \hii\ regions.  In both cases
we would hope to enhance understanding of the internal physics of the
gas within the \hii\ regions, measuring as a function of position on a
region its temperature and electron density, and applying tests such as
the [\oiii]/[\oii] emission ratio to the question of density bounding. 

The apparent invariance in the luminosity of the ``glitch'' measured in
the LF does offer a possible refined secondary standard candle of high
luminosity and constant presence for use well into the Hubble flow,
independently of whether the interpretation we have placed on it in the
present article is fully, or even partially, valid.  To demonstrate its
scope will require amplification of the number of objects studied to
determine reliably, using local galaxies, the scatter in the observed
feature, including any possible dependence on galaxy luminosity and
type, and to calibrate it using galaxies whose Cepheid distances have
been determined in the {\it HST\/} key project.

\acknowledgements

The William Herschel Telescope is operated on the island of La Palma by
the Royal Greenwich Observatory in the Spanish Observatorio del Roque de
los Muchachos of the Instituto de Astrof\'\i sica de Canarias.  This
work was partially supported by the Spanish DGICYT (Direcci\'on General
de Investigaci\'on Cient\'\i fica y T\'ecnica) via Grants PB91-0525 and
PB94-1107.  This research has made use of the NASA/IPAC Extragalactic
Database (NED) which is operated by the Jet Propulsion Laboratory,
California Institute of Technology, under contract with the National
Aeronautics and Space Administration.  Thanks are due to Andrew Cardwell
for a series of computational tests related to the \hii\ region
luminosity function.  We are grateful to Chris McKee, and to the
referee, Marshall McCall, for comments which have helped to improve this
paper.

\end{document}